\documentclass[11pt]{article}

\usepackage[utf8]{inputenc}
\usepackage{graphicx}
\usepackage{amscd}
\usepackage{amsmath}
\usepackage{amssymb}
\usepackage{amstext}
\usepackage{url}
\usepackage{epsfig}
\usepackage{wasysym}
\usepackage{fancyvrb}
\usepackage{alltt}
\usepackage{hhline}
\usepackage{psfrag}
\usepackage{array}
\usepackage{setspace}
\usepackage{float}
\usepackage{multirow}
\usepackage{slashbox}

\usepackage{ulem}

\usepackage{pst-tree}%permet d'écrire en couleur DANS LE PDF OU LE PS pas dans le DVI !!!!!

\usepackage{natbib}
\def\H{\mathcal{H}}

\def\Deg{^{\circ}}

\newcommand{\dpar}[2]{\frac{\partial #1}{\partial #2}}

\newcommand{\jacc}[2]{\frac{\partial^2 #1}{{\partial #2}^2}}

\newcommand{\documentdate}{21 January 2011}

\graphicspath{{/EPSF/}{Figures/}{./}}

\begin{document}

\begin{titlepage}
%\date{\today}

  \includegraphics[height=3.5cm]{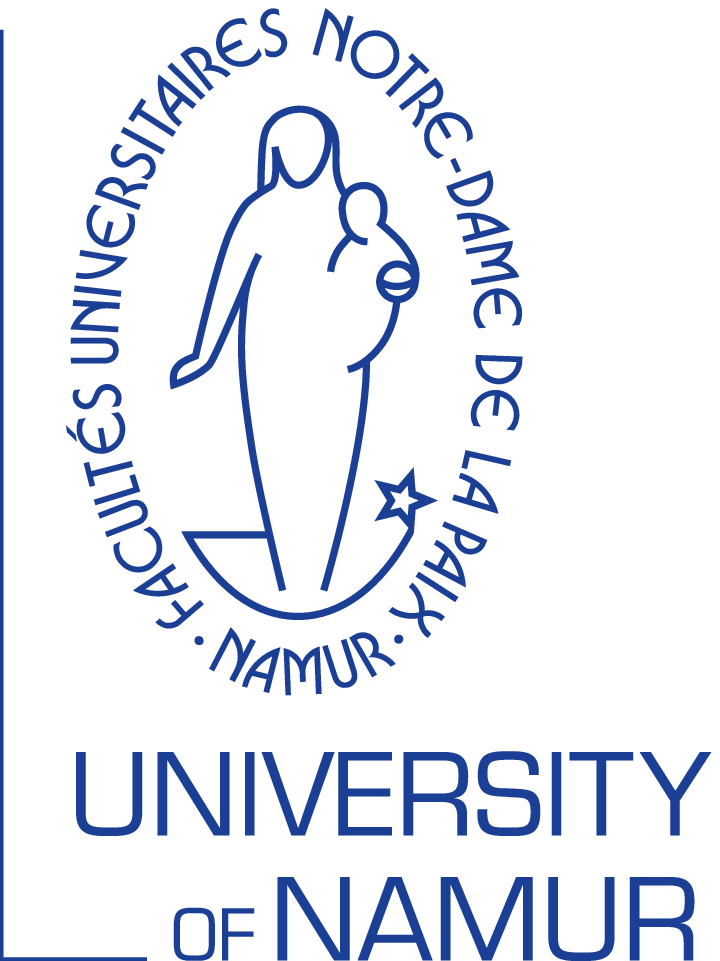}

  \vspace*{1cm}
  \hspace*{1cm}
  \fbox{\rule[-3cm]{0cm}{6cm}\begin{minipage}[c]{12cm}
      \begin{center}
        {\Large Analytical and numerical study of the ground-track resonances of Dawn orbiting Vesta}\\
        \mbox{}\\
        by Nicolas Delsate\\
        \mbox{}\\
        Report naXys-04-2011 \hspace*{20mm} \documentdate 
      \end{center}
    \end{minipage}
  }
  
  \vspace{1cm}
  \begin{center}
    \includegraphics[height=3.5cm]{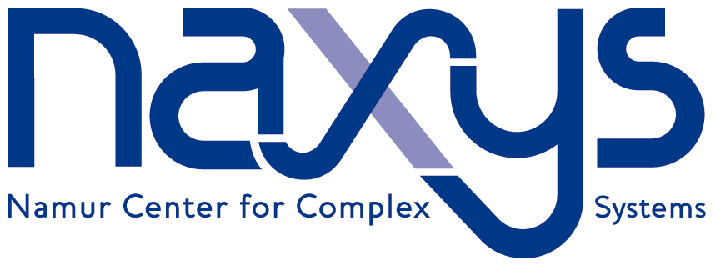}
    
    \vspace{1.cm}
           {\Large \bf Namur Center for Complex Systems}
           
           {\large
             University of Namur\\
             8, rempart de la vierge, B5000 Namur (Belgium)\\*[2ex]
             {\tt http://www.naxys.be}}
           
  \end{center}
  
\end{titlepage}

\newpage
  \title{Analytical and numerical study of the ground-track resonances of Dawn orbiting Vesta}

  \author{N. Delsate (nicolas.delsate@math.fundp.ac.be)}
  %%%%%%%%%%%%%%%%%%%%%%%%%%%%%%%%%%%%%%%%%%%%%%%%%%%%%%%%%%%%%%%%%%%%%%%%
%\maketitle
  \begin{abstract}
    %The Dawn Discovery mission was successfully launched on September 27, 
    %2007. 
    The aim of Dawn mission is the acquisition of data from orbits 
    around two bodies, (4)~Vesta and (1)~Ceres, the two 
    most massive asteroids. 

    Due to the low thrust propulsion, Dawn will slowly 
    cross and transit through ground-track resonances, 
    where the perturbations on Dawn orbit may be significant. 
    %% The ground-track resonances appear when the the Dawn's 
    %% orbital period to Vesta's rotational period ratio is one integer.
%    commensurabilities between its orbital 
%    period and Vesta's rotational period (ground-track resonances), 
    In this context, to safety go the Dawn mission from the approach orbit 
    to the lowest science orbit, it is essential to know the properties 
    of the crossed resonances. 

    This paper analytically investigates the properties of the 
    major ground-track resonances (1:1, 1:2, 2:3 and 3:2) 
    appearing for Vesta orbiters: location of the equilibria, aperture 
    of the resonances and period at the stable equilibria. 
    We develop a general method using an averaged Hamiltonian formulation 
    with a spherical harmonic approximation of the gravity field. If the 
    values of the gravity field coefficient  change, our method 
    stays correct and applicable. We also 
    discuss the effect of one error on the $C_{20}$ and $C_{22}$ coefficients 
    on the properties of the 1:1 resonance. 
    These results are checked by numerical tests. 
    %and we analytically explain 
%    the increase of the eccentricity in the resonances 1:1 and 2:3: 
%    respectively due to the $J_{32}$ and $J_{22}$ coefficients. 
    We determine that the increase of the eccentricity appearing in 
    the 2:3 resonance is due to the $J_{22}$ coefficient. 
    Finally we numerically study the probability of the capture in 
    resonance 1:1. Our paper reproduces, explains and completes the 
    results of Tricarico and Sykes (2010).\\

    {\bf Keyword}
    Vesta  -- Gravitational perturbations -- Resonance -- Spacecraft operations -- Analytical study
   
  \end{abstract}
  %%%%%%%%%%%%%%%%%%%%%%%%%%%%%%%%%%%%%%%%%%%%%%%%%%%%%%%%%%%%%%%%%%%%%%%%

%\tableofcontents

%%%%%%%%%%%%%%%%%%%%%%%%%%%%%%%%%%%%%%%%%%%%%%%%%%%%%%%%%%%%%%%%%%%%%%%%%
%%%%%%%%%%%%%%%%%%%%%%%%%%%%%%%%%%%%%%%%%%%%%%%%%%%%%%%%%%%%%%%%%%%%%%%%%
%\vspace{-1cm}
\section{Introduction}

After the explorations of the planet of our Solar System, 
the new space missions are aimed at asteroids and comets, 
moving from fly-bys to rendezvous and orbiting. 
%The new profiles of the space missions are aimed at asteroids and comets, 
%moving from fly-bys to rendezvous and orbiting. 
For these missions the space probe could have an operating orbital period 
close to the asteroid one. 
%These missions generally call for a period of orbital operations 
%in close proximity to the asteroid. 
A challenge for the navigators of 
these missions is to predict the orbital environment around the asteroid and 
to derive pre-mission plans for the control of these orbits.

Several authors have already been interested in these problems.  
\cite{Scheeres1994} studied the stability of a spacecraft in 
synchronous orbits with the asteroid rotation. \citet{Scheeres1998} 
presented the dynamics of orbits close to the asteroid 4179 Toutatis. 
\cite{Rossi1999} numerically studied the orbital evolution around 
irregular bodies with two methods. 
%with two independent codes which can integrate numerically the orbits of test 
%particles around irregularly shaped primary bodies. 
The first is based on a representation of the central body in terms 
of “mascons”, %(discrete spherical masses), 
and the second model considers the central body as a polyhedron with a 
variable number of triangular faces. 
\cite{Hu2002} investigated the spacecraft motion about a slowly rotating 
asteroid (as compared to the spacecraft orbit period) with 
the second degree and order of the gravity field. 
\cite{Lara2010} and \cite{Russell2009} were interested in the motion 
of an orbiter around Enceladus. 
More recently, \cite{Tricarico2010} studied the dynamical 
environment of Dawn around Vesta.\\

The asteroid (4) Vesta is one of the biggest minor planets of the main belt. 
Originally discovered in 1807 by Heinrich Olbers, it is one of the 
targets of the space 
mission Dawn \citep{Rayman2006,Russell2007}, that will perform three 
different altitude science orbits  before flying to Ceres. 
Dawn was successfully launched on September 27, 2007 and will arrive to Vesta 
in July 2011. In the preparation of this mission, \cite{Tricarico2010} 
proposed a numerical survey of the orbits around Vesta and 
emphasized particularly the problem of the 1:1 ground-track resonance. 
%% Due to the specific propulsion (low thrust), Dawn will be slowly 
%% transiting commensurabilities between its orbital period and Vesta's 
%% rotational period (ground-track resonances), where perturbations on 
%% Dawn's orbit may be significant. During this transit, Dawn can be 
%% cross the separatrix of the resonances and the orbit could become 
%% chaotic, or Dawn can be trapping in one resonance.
%% We are interested to go into the results of \cite{Tricarico2010} 
%% in depth. 

We revisit their study in adding a more complete numerical 
exploration and analytical explanations of the observed phenomena 
i.e. location, aperture and period of the stable equilibria of the 
ground-track resonances. 
%% Then aim of our paper is 
%% %to go into the results of \cite{Tricarico2010} in depth and 
%% to analytically study the properties of the main 
%% ground-track resonances: location, aperture and period at the 
%% stable equilibria. 
The interest of an analytical investigation is the easiness 
and the fastness to refind the results when the data change. 

Our paper is organized as follows. 
In section~\ref{secMainPert} we briefly recall the specification of 
the Dawn mission and we determine the main forces acting on 
Dawn orbiting Vesta. Then we select the most important perturbations 
that we use in our analytical models. 
In section~\ref{secNumStudy} we rebuild the results of 
\cite{Tricarico2010} and we numerically localize the ground-track 
resonances (the commensurabilities between the orbital period of the probe 
and the rotational period of the asteroid); we measure the aperture 
of these resonances and we determine 
the period at the stable equilibrium of the 1:1 resonance. 
In section~\ref{sectAnalyt} we recover the results of 
section~\ref{secNumStudy} but in a purely analytical way. We develop a 
general method using an averaged 
Hamiltonian formulation with a spherical harmonic approximation 
of the gravity field. We also discuss the effect of an error on the 
$C_{20}$ and $C_{22}$ coefficients on the properties of the 1:1 resonance. 
In section~\ref{secEgd} we 
%analytically explain the increase of the 
%eccentricity in the resonances 1:1 and 2:3.
show that the increase of the eccentricity in the 2:3 resonance is 
due to the $J_{22}$ coefficient. 
In section~\ref{secThrust} we study numerically the thrust 
and the probability of capture in the 1:1 resonance and 
we detail the results of Tricarico about the trapping. 
Finally we summarize and draw our conclusions.

\section{Main perturbations acting on Dawn around Vesta}\label{secMainPert}

\subsection{Dawn mission}
Dawn will be the first mission to orbit two main 
belt asteroids: Vesta and Ceres. They are the two most massive asteroids 
in the main belt and they are particularly interesting because they span 
the region of space between the rocky inner body Solar System and the 
wetter bodies of the outer Solar System. They have survived intact 
from the earliest days of 
the Solar System. Hence they can provide the opportunity for acting 
as windows to the conditions present in the first few millions years 
of the Solar System existence. 

For a complete overview of the Dawn mission, we refer to 
\cite{Rayman2006} and \cite{Russell2007}. 
The science observations will be concentrated in three campaigns 
around Vesta, each one conducted in a different circular, 
Sun-synchronous quasi-polar orbit. 
The first science orbit, Vesta Science Orbit 1 (VSO1), has a radial 
distance of $2\,700$~km (corresponding to an equatorial altitude 
of about $2\,400$~km) with a period of 
58~hours. Upon completion of VSO1, the thrusting will resume for the 3-week 
transfer to Vesta Science Orbit 2 (VSO2). The VSO2 distance is $950$~km 
(an equatorial altitude about $670$~km), where the orbital period is 12~hours. 
This part of the mission is named HAMO, namely High Altitude Mapping 
Orbit. The transfer from VSO2 to VSO3 will take $30$~days. At a radial 
distance of $460$~km (equatorial altitude of about $180$~km), VSO3 
is the lowest orbit planned and its orbital period is 4~hours. 
This last scientific orbit is named LAMO for Low Altitude Mapping Orbit. 
The transition between the different phases are made by a Solar 
electric propulsion with low thrust.

\subsection{Main perturbations}
The major perturbations acting on a probe orbiting an asteroid 
of the main belt are due to the discrepancies of the asteroid 
gravitational field, to the gravitational perturbations induced 
by the Sun, Mars and Jupiter as well as to the direct radiation pressure 
of the Sun. We use the following simplified relations 
\citep{Montenbruck2000} to characterize the norms of the accelerations 
induced by these 3 forces: 
\begin{eqnarray}
  a_{nm} & = & (n+1) \frac{\mu}{r^2} \frac{R_e^n}{r^n} \sqrt{\bar{C}_{nm}^2+\bar{S}_{nm}^2}\,, \nonumber\\
  a_{_{3b}} & = & \frac{2\mu_{_{3b}}}{d^3_{_{3b}}} r\,, \label{mainpert}\\
  a_{rp} & = & C_r \frac{A}{m} P_{\odot}\nonumber
\end{eqnarray}
where $a_{nm}$ is proportional to 
the coefficient of degree $n$ and order $m$ of the spherical 
harmonic development; $a_{_{3b}}$ corresponds to the acceleration 
induced by the third body (Sun, Mars or Jupiter) and $a_{rp}$ is the 
acceleration due to the direct 
radiation pressure. $R_e$ is the equatorial radius of the asteroid, 
$\bar{C}_{nm}$ and $\bar{S}_{nm}$ are the normalized coefficients 
of the gravitational potential; 
$\mu$ and $\mu_{_{3b}}$ represent the gravitational 
constants of the asteroid (Vesta or Ceres) and of the third body respectively. 
$P_{\odot}= 4.56 \times 10^{-6} \text{N/m}^2$ is the radiation 
pressure for an object located at a distance of 1~AU from the Sun 
and $C_r$ is the reflectivity coefficient appearing in the direct 
radiation pressure acceleration. 
$A/m$ is the area-to-mass ratio, $d_{_{3b}}$ is the distance between 
the object and the third body and $r$ is the distance 
between the probe and the central body (asteroid). 
The $A/m$ of Dawn is approximately equal to $0.04\;\text{m}^2$/kg 
\citep{Tricarico2010}. The values of the coefficients 
$\bar{C}_{nm}, \bar{S}_{nm}$ 
that we use can be found in Table~1 of \citet{Tricarico2010} as well 
as the values of $a_{3b}=2.36$~UA and $R_e=300$~km.
\begin{figure}[htbp] 
  \begin{center}
    \includegraphics[draft=false,width=\textwidth]{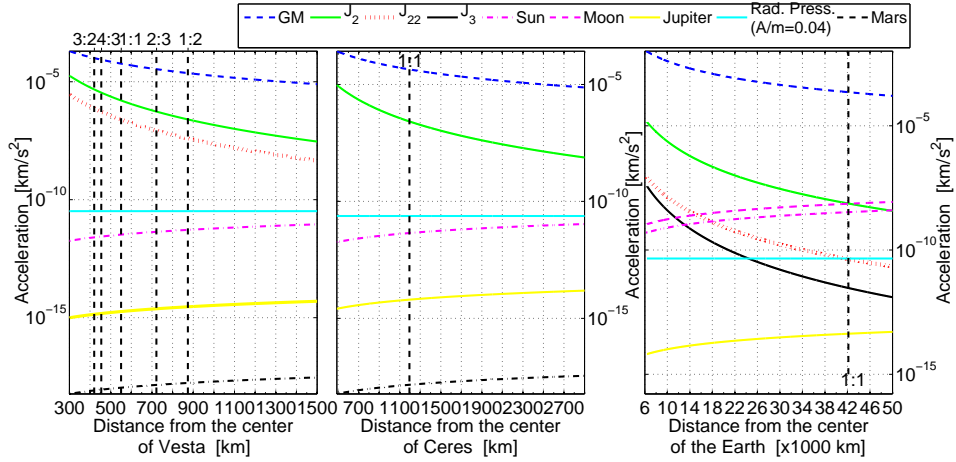}
    \caption{\label{forcesVCE} The orders of magnitude of the main 
      perturbations acting on a probe around Vesta and Ceres (left and 
      central panel) and, comparatively, around the Earth. The dashed 
      vertical lines show the ``geo''stationary and ``asterostationary'' 
      semi-major axis.}
  \end{center}
\end{figure} 

We draw the order of magnitude of the main perturbations 
\eqref{mainpert} as functions of the 
``planetocentric'' satellite distance $r$ in Fig.~\ref{forcesVCE}. 
As a consequence of the location ($\sim 2.5$~UA) 
and the shape of the asteroid (Vesta and Ceres) the order of magnitude of 
the perturbations is relatively different from the magnitude of the 
perturbations for a satellite of the Earth. For a space probe orbiting 
Ceres or Vesta, the perturbation due to the discrepancies of the 
gravitational field is several orders of magnitude greater than these due to 
the third body (or to the radiation pressure) perturbations. 
This remains true for large values of the probe semi-major axis. 
On the contrary, in the case of a stationary satellite of the Earth (GEO), 
the Luni-Solar perturbations are comparable in size to the 
Earth oblateness ones. 

Therefore, later in this paper, we shall only keep the perturbations 
due to the gravitational field of the asteroid. 

\section{Vesta dynamical environment: numerical study}\label{secNumStudy}
%% The effects due to a non-spherical asteroid shape are quite complex. 
%% To modelize the asteroid gravitational field, the spherical harmonic 
%% expansion is generally used. However  
%% %Although the approach of the spherical harmonic expansion of the 
%% %asteroid gravitational field is generally used, 
%% it is not enlightening for pre-mission planning purposes when there 
%% is no definitive estimate for the gravitational harmonics. 
%% Then a better approach may be a used for other approximations 
%% of the asteroid shape: the mascon \citep{Rossi1999}, 
%% a tri-axial ellipsoid \citep{Scheeres1994,Compere2011}, 
%% a collection of point mass potentials, a polyhedron 
%% approximation \citep{Rossi1999}\dots 
%% However the major gravitational perturbations of the asteroid shape onto 
%% the orbit of an orbiter can usually be characterized by 
%% using the spherical harmonic representation. For example, to 
%% characterize the main effect of the asteroid shape, it is 
%% convenient to use the primary oblateness term $C_{20}$. 

The effects due to a non-spherical asteroid shape are quite complex.  
Several authors gave different solutions to modelize the asteroid 
gravitational field. \citet{Rossi1999} used the mascons or the polyhedrons 
approximations. \citet{Scheeres1994} and \citet{Compere2011} 
used a tri-axial ellipsoid representation. 
Let us note that the spherical harmonic representation of the 
gravity field is uniformly convergent for $a<c\sqrt{2}$ 
\citep{Balmino1994,Pick1973} where 
$a$ and $c$ are the biggest and smallest semi-axis of the asteroid. 
This is the case of Vesta. 
Then the major gravitational perturbations of the asteroid shape 
onto the orbit of an orbiter can usually be characterized by the 
spherical harmonics. For example, to 
characterize the main effect of the asteroid shape, it is 
convenient to use the primary oblateness term $C_{20}$. 

Furthermore, using a spherical harmonic representation, 
\cite{Tricarico2010} tested several models for the mass 
density distribution of Vesta. They concluded that Dawn dynamics 
depends primarly on Vesta shape and only secondarily on the details of 
the interior structure. 
Therefore we decide to use the uniform mass density scenario of Tricarico 
with a spherical harmonic development of the gravity field.

As previously stated, the dynamics of a spacecraft orbiting 
within $\sim 1\,000$~km from Vesta is dominated by the gravitational 
field of Vesta. As a consequence the system of differential 
equations describing the probe motion is given by 
$\ddot{\boldsymbol{r}} = -\nabla U_{\text{pot}}$, 
where $\ddot{\boldsymbol{r}}_{\text{pot}}$ represents the acceleration 
induced by Vesta gravity field and can be expressed as the 
gradient of the following potential \citep{Kaula1966}
\begin{equation}\label{Vestapotential}
  U_{\text{pot}}(r,\delta,\phi) = -\frac{\mu}{r}+ \frac{\mu}{r}\sum_{n=2}^{\infty}
  \sum_{m=0}^{n} \left(\frac{R_e}{r}\right)^n\,\mathcal{P}_n^m(\sin\phi)(C_{nm}
  \cos\, m\delta+ S_{nm}\sin\, m\delta)\,, 
\end{equation}
where the quantities $C_{nm}$ and $S_{nm}$ are the spherical harmonics
coefficients of the potential of Vesta. $n$ and $m$ are respectively 
the degree and the order of the development. $\mu$ and $R_e$ are 
respectively the gravitational constant and equatorial radius of Vesta. 
The quantities $(r,\delta,\phi)$ describe the planetocentric 
spherical coordinates of the space probe. $\mathcal{P}_n^m$ are 
the associated Legendre polynomials. 
It is worth noting that all terms $S_{n0}$ are zero. 
The potential \eqref{Vestapotential} may be expressed using 
an alternative way, with a cosine terms, introducing a phase difference 
$\delta_{nm}$ as well as a coefficient $J_{nm}$:
\begin{equation}
  U_{\text{pot}}(r,\delta,\phi) = -\frac{\mu}{r}- \frac{\mu}{r}\sum_{n=2}^{\infty}
  \sum_{m=0}^{n} \left(\frac{R_e}{r}\right)^n\,\mathcal{P}_n^m(\sin\phi)(J_{nm}
  \cos(m\delta-m\delta_{nm})\,,
\end{equation}
where
\begin{equation}\label{defJnm}
  C_{nm}=-J_{nm}\cos(m\delta_{nm}), \;\; S_{nm}=-J_{nm}\sin(m\delta_{nm}), \;\; 
  m\delta_{nm}=\arctan\left(\frac{-S_{nm}}{-C_{nm}}\right).
\end{equation}
\cite{Tricarico2010} have shown that the dynamical behavior of Dawn 
is dominated by the first 8 degrees of the gravitational potential and that the 
main resonant terms are of degree 3 and 4. Henceforth we model 
the gravity potential of the central body using the coefficients 
until degree and order 4. The coefficients used here can be found in the 
Table~1 of \citet{Tricarico2010}. 
%We adopt the variable step size Bulirsch-Stoer 
%algorithm (see e.g. \citealt{bulirsh-stoer}) to numerically integrate 
%the differential equations. Let us note that, for the purpose 
%of validation, we also use a second numerical integrator 
%Adams-Bashforth-Moulton $10^{th}$ order predictor-corrector.

The numerical integrations of the differential equations 
are made by a home-made numerical software: 
{\bf N}umerical {\bf I}ntegration of the {\bf M}otion of an 
{\bf A}rtificial {\bf S}atellite orbiting a {\bf TE}lluric {\bf P}lanet,  
for short \verb|NIMASTEP|. This extensive tool allows to derive the osculating 
motion of an arbitrary object orbiting any of the terrestrial planets 
of our Solar System. This software has been elaborated by N. Delsate 
and successfully used in \cite{Valk2009b} and \cite{Delsate2010}. 
It has been recently extended for asteroids in \cite{Compere2011}.
%the special needs of the investigations developed in this 
%work and it is used in an other 

%\subsection{Maps of variation of semi-major axis, eccentricity and radius}\label{sectMap}
\subsection{Maps of variation of semi-major axis, inclination and radial distance}\label{sectMap}
\begin{figure}[htbp] 
  \begin{center}
    \includegraphics[draft=false,width=\textwidth]{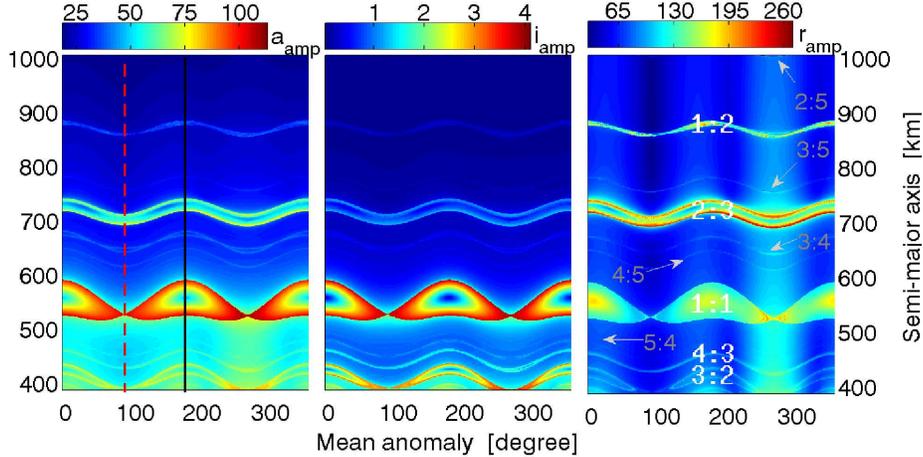}
    \caption{\label{DgaExcRayDawn}
      The semi-major axis, the %eccentricity and the distance computed as a 
      inclination and the radial distance computed as 
      functions of the initial mean anomaly and the initial 
      semi-major axis. The equations of motion include the central 
      body attraction, the harmonics until degree and order 4. 
      The anomaly step is $1\Deg$ and the semi-major axis step is $700$~m. 
      The initial conditions are $i_0 = 90\Deg$, $\Omega=\omega=0\Deg$ 
      and $\theta=0\Deg$. The integration time is 1 year. 
      %from epoch fixed at 14 September 2019. 
      The patterns have been obtained by plotting the amplitude of variation 
      of the semi-major axis ($\text{a}_{\text{amp}}$) in the left panel, the 
      amplitude of variation of the %eccentricity ($\text{e}_{\text{amp}}$) 
      inclination ($\text{e}_{\text{amp}}$) 
      in the central panel and the amplitude of variation of 
      the radial distance in the right panel ($\text{r}_{\text{amp}}$). 
      These amplitude are shown by the colorbars.}
  \end{center}
\end{figure}

In Fig.~\ref{DgaExcRayDawn}, we plot the results 
of the numerical integrations performed for a 
set of $316\,800$ orbits, propagated over a time span of 1~year 
with a entry-level step size of 100~seconds. We consider a set of initial 
conditions defined by a mean anomaly grid of $1\Deg$ and a
semi-major axis grid of 700~m, spanning the $[385, 1\,001]$~km range. 
In this work, we explore the dynamics of the spacecraft Dawn in a polar 
orbit ($i_0 = 90\Deg$) because the initial inclination of Vesta is 
Sun-synchronous, thus corresponds to a quasi-polar orbit. 
The other fixed initial conditions are $\Omega=\omega=0\Deg$ for 
the longitude of the ascending node and the 
argument of pericenter respectively, and $\theta=0\Deg$ for the sideral time. 
%at epoch 14 September 2019.

We show the amplitude of the semi-major axis (namely the difference 
between the maximum and minimum semi-major axis reached during the 
integration), %eccentricity and distance of each orbit respectively 
inclination and radial distance of each orbit respectively 
in the left, central and right panels of Fig.~\ref{DgaExcRayDawn}. 
These figures allow to locate the 
{\it ground-track} resonances. Indeed, when the orbital period of the 
space probe is close to a commensurability with the rotational period 
of Vesta, the repeating ground-track of the probe will periodically 
encounter the same configuration of the gravitational field. 
This is what we call a {\it ground-track resonance}. 
Then a part of the effects of the gravitational perturbations is amplified 
leading to a motion with very long periods 
and large perturbations in the semi-major axis %and eccentricity 
of the spacecraft orbit. This is what we observe in the left panel 
%and middle panels 
of Fig.~\ref{DgaExcRayDawn} where the variation of the 
semi-major axis %and of the eccentricity are 
is clearly largest at $\sim 550$~km and $\sim 720$~km, as shown by 
the color code.

It is obvious that the usual pendulum-like phase spaces 
\citep{Gedeon1969} (appearing in the case of a ground-track resonance) 
are distorted like a wave. 
Actually, it is worth noting that, at first, the sampling is 
carried out with respect to the osculating initial conditions.
Second, within the framework of mean motion theory, it 
is well-known \citep{Exertier1995} that, due to the short-period 
oscillations, the mean and the osculating initial conditions cannot 
be considered as equal. In other words, for the same fixed value
of the initial osculating semi-major axis and for various initial
mean anomalies, we obtain different values for the mean
semi-major axis. For more explanation, in a geostationary 
satellite case, the reader can refer to \cite{Valk2009b}. \\

The results obtained here are similar to \citet{Tricarico2010} ones 
(Fig.4 in their article). 
We easily find 5 main ground-track resonances (noted in white in 
Fig.~\ref{DgaExcRayDawn}): 3:2, 4:3, 1:1, 2:3 and 1:2. 
It is also possible to find other resonances (5:4, 4:5, 3:4, 5:3 and 
5:2) but they are very thin (noted in gray with an arrow in 
Fig.~\ref{DgaExcRayDawn}). 
The amplitude of variation of the inclination is near $1\Deg$, 
except around the separatrix of the 1:1 resonance where the amplitude 
of variation reaches $3.8\Deg$. In consequence, in the future, 
we shall consider the inclinations as equal to $90\Deg$. 
In the right panel of Fig.~\ref{DgaExcRayDawn}, 
we notice that the two most important resonances are the 1:1 and the 2:3 
resonances: the largest is the 1:1 but the strongest perturbations 
on the radial distance appear in the 2:3 resonance.  
This last observation will be explained in this paper (Section~\ref{secEgd}). 

\subsection{Location and aperture of the main ground-track resonances}
Now let us determine the location and the aperture of the 
main resonances. We perform two initial mean anomaly sections 
(Fig.~\ref{DgaExcRayDawn}) near a stable (vertical 
continued black line) and an unstable (dashed red line) equilibrium 
point of the resonance 1:1. 
\begin{figure}[htbp] 
  \begin{center}
    \includegraphics[draft=false,width=\textwidth]{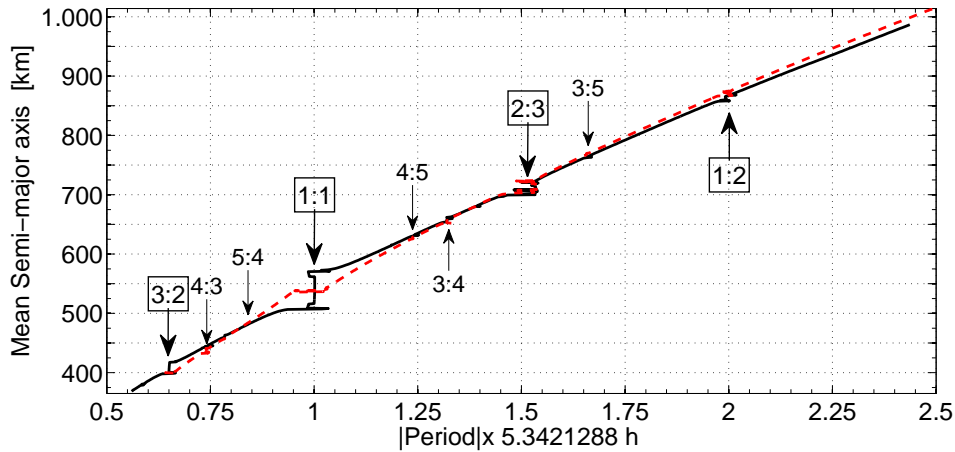}
    \caption{\label{PeriodTranche}
      The mean semi-major axis of a probe around Vesta with respect to 
      its mean anomaly period. The periods are obtained by frequency 
      analysis (NAFF). The estimation of the periods is made over a 
      period of 1 year. The model used is the same as in 
      Fig.~\ref{DgaExcRayDawn}. 
      The continued black and the dashed red lines correspond to the vertical 
      lines of Fig.~\ref{DgaExcRayDawn} (left panel). 
      We tick off the detectable ground-track resonances.}
  \end{center}
\end{figure} 
%We perform two mean anomaly sections, the vertical 
%continued black line and the dashed red line in the left panel 
%of Fig.~\ref{DgaExcRayDawn}, that are the sets of orbits having 
%the same initial mean anomaly value respectively near a stable 
%and an unstable equilibrium point of the resonance 1:1. 
For each 
orbit of these two sections, we compute, using the Numerical 
Analysis of Fundamental Frequencies, for short 
NAFF \citep{Laskar1988,Laskar2005}, the fundamental period of 
the mean anomaly with respect to the mean initial semi-major axis $a_0$. 
The mean semi-major axis is calculated with a numerical averaging 
method applied on the osculating semi-major axis, on 5 revolutions 
of Dawn around Vesta. 
The results are presented in Fig.~\ref{PeriodTranche} and in 
Tab.~\ref{TabLocalThcikReso}. For more explanations 
and other applications of the using of NAFF in space geodesy we refer to 
\cite{Lemaitre2009}.

\begin{table}[htbp]
  \begin{center}
    \caption{The mean semi-major axis location and the aperture of the 
      main resonances (using Fig.~\ref{PeriodTranche}).}
    \label{TabLocalThcikReso}
    \begin{tabular}{l | >{$}c<{$} >{$}c<{$} >{$}c<{$} >{$}c<{$} }
      \hline \hline
      Resonance & \text{3:2} & \text{1:1} & \text{2:3} & \text{1:2}\\
      \hline
      Mean Semi-major axis  [km] & 409.1 & 539.6 & 717.3 & 871.5\\
      Aperture  [km] & 18 & 63 & 22.5 & 7.5
    \end{tabular}
  \end{center}
\end{table}

\subsection{Period of the resonant angle at the stable equilibrium of the 1:1 resonance}
\begin{figure}[htbp] 
  \begin{center}
    \includegraphics[draft=false,angle=-90,width=0.5\textwidth]{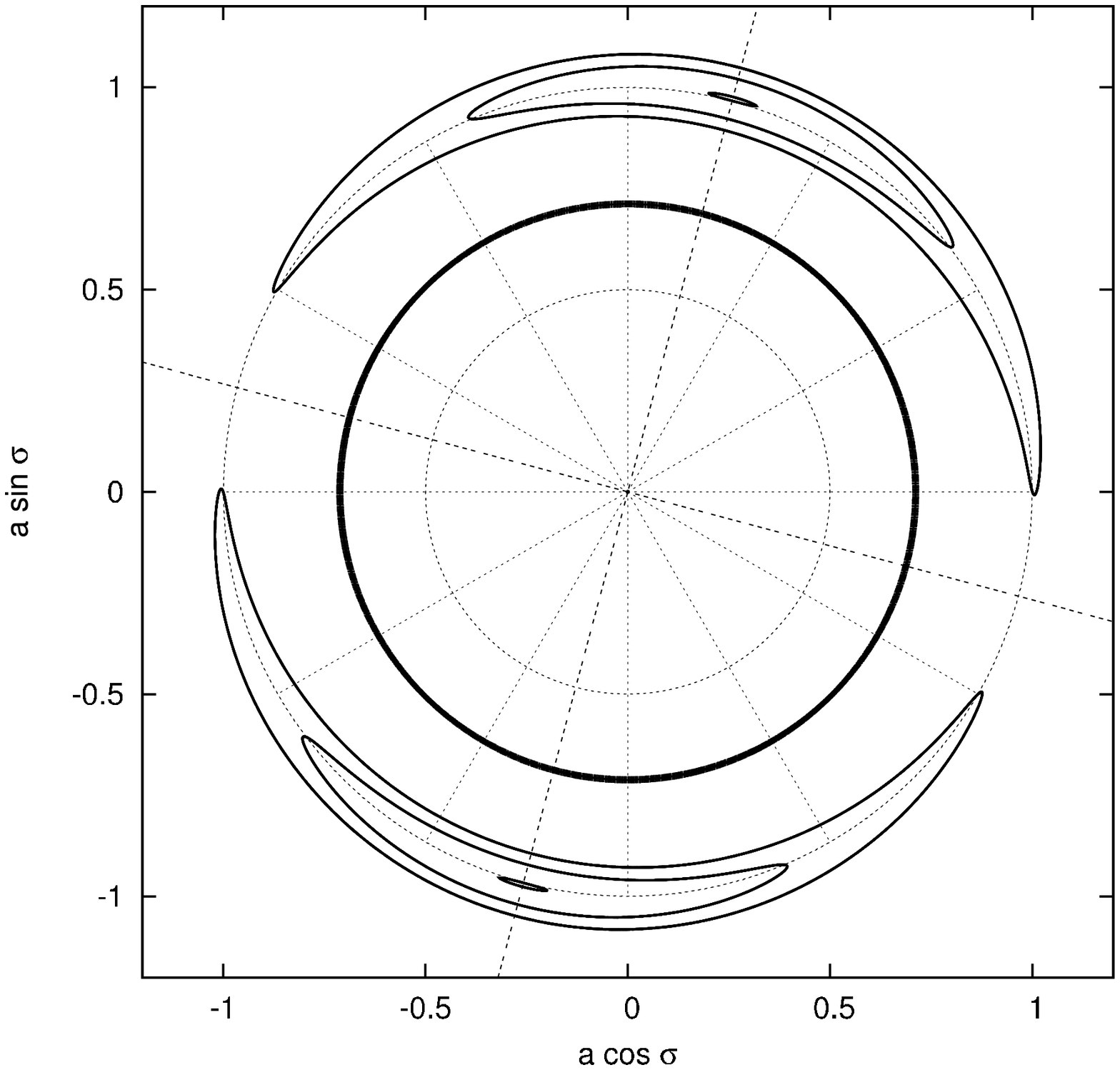}
    \caption{\label{C22PhaseSpace} General illustration of the motion of a 1:1 
      ground-track resonant probe near stable equilibria in a rotating 
      reference frame as seen from the pole. The variations on the mean 
      semi-major axis have been amplified for the illustration.}
  \end{center}
\end{figure}

In the case of a 1:1 ground-track resonance, the resonant 
angle is defined by $\sigma=\lambda-\theta=(\Omega+\omega+M)-\theta$ 
where $\theta$ is the sideral time \citep{Gedeon1969,Valk2009a}, 
$\lambda$ is the mean longitude, $\Omega$ is the ascending node, 
$\omega$ is the argument of pericenter and $M$ is the mean anomaly. 
For illustration, we draw in Fig.~\ref{C22PhaseSpace} the motion 
of a probe in resonance 1:1 with its central body. 

We can determine the period of the resonant angle $\sigma$ at 
the center of the banana curves. For that, 
we use the {\bf N}amur {\bf A}lgorithm {\bf F}or 
{\bf F}orced {\bf O}scillations, for short 
\verb|NAFFO| \citep{Noyelles2011}, to determine numerically the 
semi-major axis and the value of the resonant angle corresponding to the 
stable equilibrium. For clarity, the algorithm is reminded 
here: the initial conditions ($a,\sigma$) are conveniently chosen 
to be close enough to the equilibrium, here for example $a=550$~km 
and $\sigma=90\Deg$. Now we try to minimize the free libration 
(the oscillation of $\sigma$ and $a$ in the banana curves). 
This minimization is performed iteratively by 
\begin{enumerate}
\item[(1)] performing a numerical integration of the equations of the motion;
\item[(2)] identifying the free component by a frequency analysis 
  (here using NAFF);
\item[(3)] removing the free component from the initial conditions 
  to determine a new initial condition. Then a new
  numerical integration is being performed (return to step 1).
\end{enumerate}
We apply this algorithm to different models containing only 
the gravity field of the central body. They respectively contain 
the $J_{22}$ term ($J_{nm}$ are defined by Eq.~\eqref{defJnm}), 
the $J_{22}$ and $J_2$ terms ($-C_{20}=J_{20}\stackrel{not.}{=}J_2$), 
and finally the whole gravity field until degree and order 4 ($n=4=m$). 
When the algorithm stops, we make a frequency analysis 
(NAFF) of the signal ($a,\sigma$) to determine the period of the 
resonant angle at the equilibrium. 

%%%%%
%%%%%PR DANS LA THESE
%%%%%
%% In Fig.~\ref{FigNAFFO} we present 
%% the result of the two iterations of NAFFO for the $J_2+J_{22}$ case. 
%% The shifted center of libration ($\sim 1\deg$) and the non-null variation of 
%% the semi-major axis are due to the full model used containing higher order 
%% of spherical harmonic and inducing an osculating motion of the center 
%% of libration.
%% \begin{figure}[htbp] 
%%   \begin{center}
%%     \includegraphics[draft=false,width=0.85\textwidth]{VestaJ44NAFFOevol.eps}
%%     \caption{\label{FigNAFFO} Numerical integration of the motion 
%%       of a probe around Vesta with the $J_2$ and $J_{22}$ terms of the 
%%       gravity filed. Each integration correspond to one step of NAFFO.}
%%   \end{center}
%% \end{figure} 

The results of NAFFO are summarized in Tab.~\ref{TabPriodDgaNum11}. 
%for each model of force. 
Let us note that we calculate an average of the resonant 
angle and of the semi-major axis with respect to the time 
to obtain a {\it mean} value. 
The $11.21$~km difference between the $J_{22}$ and the $J_{22}+J_{2}$ model 
underlines the fact that the $J_2$ term induces an important shift 
on the semi-major axis location. 
%It is worth noting that the $J_2$ term induces a shift of 
% on the semi-major axis location. 
\begin{table}[htb]
  \begin{center}
    \caption{Numerical study of the location of the 1:1 resonant for 
      different models. The first two lines are given by the NAFFO 
      algorithm. The {\it mean} semi-major axis and the {\it mean} 
      resonant angle are obtained by numerical average with respect 
      to the time. The third line is obtained using NAFF.}
    \label{TabPriodDgaNum11}
    \begin{tabular}{l | >{$}c<{$} >{$}c<{$} >{$}c<{$} }
      \hline \hline
      Model & J_{22} & J_{2}+J_{22} & \text{All until } n=m=4\\
      \hline
      {\it Mean} semi-major axis (km) & 549.63 & 538.42 & 538.89\\
      {\it Mean} resonant angle $\sigma$ (degree) & 90\Deg & 90.814\Deg & 91.144\Deg\\
      Period (day) & 2.45 & 2.47 & 2.56
    \end{tabular}
  \end{center}
\end{table}

Our semi-major axis location of the 1:1 stable equilibria shows a good 
agreement with Fig.~4 of \cite{Tricarico2010}. For the period 
of the resonant angle, \citet{Tricarico2010} obtained a value close 
to $2.5$ days (Fig.~8 in their article) that coincides well with 
our result. Remark that, for this main resonance, our numerical 
simplified model $J_2+J_{22}$ obtains a good approximation 
of the stable equilibrium location and of the resonant angle period.

\section{Ground-track resonances: analytical study}\label{sectAnalyt}
In this section, we study analytically the main ground-track resonances 
(1:1, 1:2, 2:3 and 3:2) 
found in the previous section. First, we present a general Hamiltonian 
describing the motion of a probe orbiting an irregular central body. Then, 
we identify the main ground-track resonances. For each of them we 
simplify the Hamiltonian in the circular polar case, we explain the 
resonant angle, we determine the location and the aperture of the 
resonance and we calculate the period of the resonant angle at the 
stable equilibrium. The results are compared with the 
previous numerical results. \\
For the 1:1 resonance we also study the circular equatorial case 
for comparison and we investigate the effect of an error on $C_{20}$ 
and $C_{22}$ on the properties of the resonance.

\subsection{Hamiltonian in terms of the orbital elements and condition for ground-track resonances}
%% To describe motion about the asteroid, we write the 
%% equations of motion of a space probe in a body-fixed coordinate system. 
%% We taking account only the gravitational effect of the central body and 
%% we neglect the other force as point out above.

%\subsubsection{The Hamiltonian}
%% The asteroid's gravity potential is given by 
%% \begin{equation}\label{gravitypotential}
%%   U(r,\lambda,\phi) = -\frac{\mu}{r}+ \frac{\mu}{r}\sum_{n=2}^{\infty}
%%   \sum_{m=0}^{n} \left(\frac{R_e}{r}\right)^n\,\mathcal{P}_n^m(\sin\phi)(C_{nm}
%%   \cos\, m\lambda+ S_{nm}\sin\, m\lambda)\,, 
%% \end{equation}
%% where the quantities $C_{nm}$ and $S_{nm}$ are the spherical harmonics
%% coefficients of the gravity field. $\mu$ and $R_e$ are respectively 
%% the gravitational constant and equatorial radius of the asteroid. 
%% The quantities $(r,\lambda,\phi)$ are the planetocentric  
%% spherical coordinates of the space probes. $\mathcal{P}_n^m$ are 
%% the associated Legendre polynomials. 
%% It can be shown easily that the coefficients $C_{10}$, $C_{11}$ and $S_{11}$ 
%% correspond to the center of mass coordinates divided by the equatorial 
%% asteroid’s radius. Therefore, these coefficients are zero if the coordinate 
%% system refers to the asteroid’s center of mass. Similarly, the coefficients 
%% $C_{21}$ and $S_{21}$ are zero as long as the z-axis is aligned with the 
%% asteroid’s main axis of inertia. Moreover, it is worth noting that all
%% terms $S_{n0}$ are zero.

The potential \eqref{Vestapotential} may be expressed 
using an alternative way, with the satellite orbital elements 
($a,e,i,\Omega,\omega,M$) and the sideral time $\theta$, 
\cite{Kaula1966} represents the gravitational field as 
\begin{equation}
  U=-\frac{\mu}{r}+\sum_{n=2}^{\infty}\sum_{m=0}^{n}\sum_{p=0}^{n}\sum_{q=-\infty}^{+\infty}
  \frac{\mu}{a}\left(\frac{R_e}{a}\right)^n \, F_{nmp}(i) \,G_{npq}(e)\, 
  S_{nmpq}(\Omega,\omega,M,\theta)\,,
\end{equation}
where the functions $S_{nmpq}$ depend on the gravity field coefficient 
$C_{nm}$ and $S_{nm}$, 
\begin{equation}\label{Snmpq}
  \begin{array}{lcl}
    S_{nmpq}(\Omega,\omega,M,\theta) & = &
    \left[\begin{array}{c}
        +C_{nm}\\
        -S_{nm}
      \end{array}\right]^{n-m \text{ even}}_{n-m \text{ odd}}
    \cos \Theta_{nmpq}(\Omega,\omega,M,\theta)\\
    \\
    &+& \left[\begin{array}{c}
        +S_{nm}\\
        +C_{nm}
      \end{array}\right]^{n-m \text{ even}}_{n-m \text{ odd}}
    \sin \Theta_{nmpq}(\Omega,\omega,M,\theta)\,,
  \end{array}
\end{equation}
where $n,m,p,q$ are integers, $\theta$ is the sideral time and 
\begin{equation}
  \Theta_{nmpq}(\Omega,\omega,M,\theta)=(n-2p)\omega+(n-2p+q)M+m(\Omega-\theta)
\end{equation}
is Kaula's gravitational argument. 

$F_{nmp}(i)$ and $G_{npq}(e)$ are respectively the inclination 
and eccentricity functions. To have a complete list of the values of 
these functions, we refer to \cite{Kaula1966} and \cite{Chao2005}. 
The third index q of the eccentricity functions determines the lowest 
power of the eccentricity appearing in the potential.

With the momentum $L=\sqrt{\mu\,a}$ conjugated to $\lambda=\Omega+\omega+M$
%With $L=\sqrt{\mu\,a}$, 
the Hamiltonian describing the motion of a probe 
around an irregular central body is given by 
\begin{equation}\label{HamGravi}
  \H=-\frac{\mu^2}{2L^2}+ \sum_{n=2}^{\infty}\sum_{m=0}^{n}\sum_{p=0}^{n}
  \sum_{q=-\infty}^{+\infty} R_e^{n}\,\frac{\mu^{n+2}}{L^{2n+2}}
  \,F_{nmp}(i)\,G_{npq}(e)\, S_{nmpq}(\Omega,\omega,M,\theta)+\dot{\theta}\Lambda\,.
\end{equation}
$\dot{\theta}\Lambda$ is due to the asteroid rotation, the 
quantity $\Lambda$ is the momentum conjugate to the sidereal time $\theta$.\\
%In a resonant region 
%($\exists\, q_1,q_2\in\N \,|\, q_1\dot{M}=q_2 \dot{\theta}$) the period of 
%revolution of the probe around the asteroid is proportional to the 
%rotation time (spin) of the asteroid. In this case, the Hamiltonian 
%\eqref{HamGravi} could be restricted to et moyennée !!!!!
%\subsubsection{Ground-Track Resonances}
%% The rotation period of a probe in asteroid’s orbit is said 
%% to be in resonance if a small integer number $q_1$ of sidereal days 
%% ($\dot{\theta}$) is equal to a small integer number $q_2$ of revolution 
%% periods ($\dot{M}$) of the object, that is
%% \begin{equation}
%%   q_2\dot{M}=q_1 \dot{\theta}
%% \end{equation}
%% These resonances 

The ground-track resonances occur when the rate of Kaula 
gravitational argument is close to zero, that is
\begin{equation}{\label{resoCond}}
  \dot{\Theta}_{nmpq}(\dot{\Omega},\dot{\omega},\dot{M},\dot{\theta})=(n-2p)\dot{\omega}+(n-2p+q)\dot{M}+m(\dot{\Omega}-\dot{\theta})\simeq 0
\end{equation}
%% Typically, when the condition q = 0 is satisfied, that is when 
%% considering a zero-order expansion with respect to the eccentricity, we have
%% \begin{equation}{\label{resoCondq0}}
%%   (n-2p) ( \dot{\omega} + \dot{M} ) \simeq m(\dot{\theta} - \dot{\Omega})
%% \end{equation}
where p and q are small integer numbers. 

\subsection{Resonance 1:1}
A similar (more detailed) work has been performed in \cite{Valk2009a}. 
In this paper the authors studied the geostationary space debris 
($e=0$, $i=0\Deg$) without taking into account the $J_2$ term. 

It is well know \citep{Gedeon1969,Scheeres1994,Valk2009a} 
that the resonance 1:1 is due to the second degree and 
order of the gravity field. From Eq.~\eqref{HamGravi}, let us 
write the Hamiltonian of the second order and degree harmonic, 
denoted by $\H_{1:1}$:
\begin{equation}
  \H_{1:1}=-\frac{\mu^2}{2L^2}+ \sum_{m=0}^{2}\sum_{p=0}^{2}
  \sum_{q=-\infty}^{+\infty} R_e^{2}\,\frac{\mu^4}{L^6}
  \,F_{2mp}(i)\,G_{2pq}(e)\, S_{2mpq}(\Omega,\omega,M,\theta)+\dot{\theta}\Lambda
\end{equation}
with 
$\Theta_{2mpq}=2(1-p)\omega+(2-2p+q)M+m(\Omega-\theta)$. 
We consider the z-axis as aligned with the asteroid main axis 
of inertia ($C_{21}=0=S_{21}$), $q=0$ (null eccentricity) 
and the resonant condition \eqref{resoCond} becomes: $p=1-\frac{m}{2}$;
then the Hamiltonian writes ($S_{20}=0$):
%% \begin{eqnarray}
%%   \H_{1:1}&=&-\frac{\mu^2}{2L^2}+ R_e^{2}\,\frac{\mu^4}{L^6}
%%   \left(\sum_{p=0}^{2} \,F_{20p}(i)\,G_{2p0}(e)\,C_{20} \cos(\Theta_{20p0})\right)\\
%%   &+& \left(\sum_{p=0}^{2} \,F_{22p}(i)\,G_{2p0}(e)\,\Big(C_{22}\cos(\Theta_{22p0}) + S_{22}\sin(\Theta_{22p0})\Big)\right)\nonumber
%% \end{eqnarray}
\begin{eqnarray}\label{HamJ22_1}
  \H_{1:1}&=&-\frac{\mu^2}{2L^2}+ R_e^{2}\,\frac{\mu^4}{L^6}
  \Big[F_{201}(i)\,G_{210}(e)\,C_{20} \cos(\Theta_{2010})\\
  &+& F_{220}(i)\,G_{200}(e)\,\Big(C_{22}\cos(\Theta_{2200}) + S_{22}\sin(\Theta_{2200})\Big)\Big]+\dot{\theta}\Lambda\nonumber
\end{eqnarray}
where $\Theta_{nmpq}, F_{nmp}(i)$ and $G_{npq}(e)$ are presented in 
Tab.~\ref{TabCJ2J22}.
\begin{table}[ht]
  \begin{center}
    \caption{Expression of the functions $\Theta_{nmpq}, F_{nmp}(i)$ and
      $G_{npq}(e)$ with respect to the values of $n=2$, $m\in\{0,2\}$, 
      $p=1-m/2$ and $q=0$.}
    \label{TabCJ2J22}
    \begin{tabular}{c c c c | >{$}c<{$} | >{$}c<{$} | >{$}c<{$} }
      \hline \hline
      n & m & p & q & \Theta_{nmpq} & F_{nmp}(i) & G_{npq}(e)\\
      \hline
      2 & 0 & 1 & 0 & 0 & \frac{3}{4}\sin^2 i - \frac{1}{2} & \frac{1}{(1-e^2)^{\frac{3}{2}}}\\
      &&&&&&\\
      2 & 2 & 0 & 0 & 2\omega+2M+2(\Omega-\theta) & \frac{3}{4} (1+\cos i)^2 & (1-\frac{5}{2}e^2 + \frac{13}{16}e^4 +\dots)
    \end{tabular}
  \end{center}
\end{table}

\noindent Let us define the so-called resonant angle $\sigma$ 
\citep{Gedeon1969,Scheeres1994,Valk2009a} 
\begin{equation}
  \sigma\,=\,\lambda-\theta\,=\,(\Omega+\omega+M)-\theta\,.
\end{equation}
In order to keep a canonical set of variables, we use a %following 
symplectic transformation 
%% \begin{equation}\label{symplectictransfo}
%%   d\sigma L' + d\theta' \Lambda' = d\lambda L + d\theta \Lambda\,,
%% \end{equation}
leading to the new set of canonical variables
\begin{equation}\label{canonicalvariable}
  \sigma, \qquad L' = L, \qquad \theta' = \theta, \qquad \Lambda' = \Lambda + L.
\end{equation}
Then the Hamiltonian \eqref{HamJ22_1} becomes 
(by selecting only the resonant contributions or in other words, 
by averaging over the fast angle $\theta'$)
\begin{eqnarray}\label{HamJ22}
  \H_{1:1}&=&-\frac{\mu^2}{2L'^2}- R_e^{2}\,\frac{\mu^4}{L'^6}
  \Bigg\{ C_{20} \frac{\frac{3}{4}\sin^2 i - \frac{1}{2}}{(1-e^2)^{\frac{3}{2}}}\\
  &+& \frac{3}{4} (1+\cos i)^2 (1-\frac{5}{2}e^2 + \frac{13}{16}e^4 +\dots) \big(C_{22}\cos(2\sigma)+S_{22}\sin(2\sigma)\big)\Bigg\}
   - \dot{\theta}L' \,.\nonumber  %+ \dot{\theta}(\Lambda'-L')\,.\nonumber
\end{eqnarray}
%% Now, we average this Hamiltonian with respect to the mean motion of the 
%% probe:
%% \begin{equation}\label{EqHAverage}
%%   \big<{\H}_{J_{22}}\big>_M=\int_0^{2\pi}\H_{1:1} dM\,.
%% \end{equation}
%% For simplicity, afterwards, we miss the $\big<\; \big>_M$  
%% because the presented Hamiltonians will be always averaged. 

%% Because there is not explicit mean anomaly in the initial Hamiltonian 
%% \eqref{HamJ22}, 
%% after this averaging the remaining terms are actually equal to 
%% those that dit not explicitly contains the mean anomaly from the start. 
%% So the averaged Hamiltonian $\big<{\H}_{J_{22}}\big>_M$ \eqref{EqHAverage} 
%% is equal to $\H_{1:1}$ \eqref{HamJ22}.
$\Lambda'$ is now constant and dropped. 
During the research stages, the orbit of Dawn is assumed circular ($e=0$).
%circular. Therefore, from now, we assume $e=0$.

\subsubsection{Circular equatorial case: $e=0,\,i=0$}\label{sec11eqcas}
In the circular ($e=0$) equatorial ($i=0\Deg$) case, the 
Hamiltonian \eqref{HamJ22} becomes 
\begin{equation}\label{reso11i0}
  \H_{1:1}=-\frac{\mu^2}{2L'^2}- R_e^{2}\,\frac{\mu^4}{L'^6}
  \Bigg\{ -\frac{1}{2}C_{20} + 3 \big(C_{22}\cos(2\sigma)+S_{22}\sin(2\sigma)\big)\Bigg\} - \dot{\theta}L' \,.  %+ \dot{\theta}(\Lambda'-L')\,.
\end{equation}
Let us note that the eccentricity and the inclination stay constant because 
the Hamiltonian does not explicitly depend on the associated angles 
$\omega$ and $\Omega$.\\

\cite{Scheeres1994} studied the stability of bodies in orbits which are 
near-synchronous with the asteroid (central body) rotation. If the 
asteroid is of type I, like Vesta, then the stationary orbit contains 
two stable and two unstable equilibria. If the asteroid is of type II, 
the motion associated with near synchronous orbit is always unstable. 

\noindent Two stable equilibria\footnote{The subscrit $_s$ for {\bf s}table and $_u$ for {\bf u}nstable.} $(\sigma^*_{s1},L'^*_{s1}), (\sigma^*_{s2},L'^*_{s2})$
as well as two unstable equilibria 
$(\sigma^*_{u1},L'^*_{u1})$, $(\sigma^*_{u2},L'^*_{u2})$ are found to be 
solutions of
\begin{equation}
  \dpar{\H}{L'}=\dpar{\H}{\sigma}=0\,.
\end{equation}
The values of $\sigma^*_{s\,i}$ and $\sigma^*_{u\,i}$ are 
\begin{equation}\label{sigmaloc11}
  \sigma^*=\frac{1}{2}\arctan\left(\frac{S_{22}}{C_{22}}\right)+k\frac{\pi}{2},
  \qquad k\in\{0,1,2,3\}\,.
\end{equation}
%\begin{equation}
%  \begin{array}{rclrcl}
%    \sigma^*_{11}&=&\lambda^*&\qquad \sigma^*_{12}&=&\lambda^*+\pi\\
%    \sigma^*_{21}&=&\lambda^*+\frac{\pi}{2}&\qquad \sigma^*_{22}&=&\lambda^*+\frac{3\pi}{2}
%  \end{array}
%\end{equation}
%% where $\lambda^*$ is the first quadrant solution of
%% \begin{equation}
%%   \tan 2\lambda^* = \frac{S_{22}}{C_{22}}\,,
%% \end{equation}
This formula remains valid for any value of $C_{22}$ and $S_{22}$. 
For the particular value of 
Vesta\footnote{The corresponding normalized coefficient is equal to $\bar{C}_{22}=0.004\,771$.}, $C_{22}=3.079667257459264\times 10^{-3}$ and 
$S_{22}=0$ \citep{Tricarico2010}, we obtain:
\begin{equation}
  \sigma^*_{s1}=\pi/2,\quad \sigma^*_{s2}=3\pi/2\quad \text{and} \quad 
  \sigma^*_{s1}=0    ,\quad \sigma^*_{s2}= \pi\,.
\end{equation}
The corresponding equilibria in semi-major axis, depending on these values 
and on the value of $C_{20}$, are the solutions $L'^*$ of the equation:
\begin{equation}\label{EqEquilL7}
%  \dot{\theta}L'^7-L'^4\mu^2-\frac{3}{2}\mu^4R_e^2 (C_{20}-3C_{22}) = 0
  \dot{\theta}L'^7-L'^4\mu^2-\frac{3}{2}\mu^4R_e^2 \left( C_{20} \pm 3 \frac{C_{22}^2+S_{22}^2|C_{22}|}{\sqrt{C_{22}^2+S_{22}^2}}  \right) = 0\,.
\end{equation}
The sign $+$ or $-$ corresponds respectively to the unstable and the 
stable points. The solutions of this equation are given in Tab.~\ref{Eq11J2i0}. 
We point out that to neglect the $C_{20}$ term leads to an error 
spread from $9.5$ to $31.9$~km with a mean value of about %$\pm 4.5$~km 
$20.7$~km on the location of the equilibria i.e. a relative 
variation of $\sim 3.76\%$. For comparison, this difference is about 
500~m for a GEO i.e. a relative variation of $0.001\%$.
\begin{table}[ht]
  \begin{center}
    \caption{The circular equatorial case: semi-major axis location 
      of the equilibria (stable $a^*_{s1}=a^*_{s2}\stackrel{not.}{=}a^*_{s}$  
      and unstable $a^*_{u1}=a^*_{u2}\stackrel{not.}{=}a^*_{u}$) 
      of the 1:1 resonance (second and third columns) 
      with respect to the value of $C_{20}$ (non-normalized coefficient). 
      In fourth and fifth columns the analytical calculation of the 
      period \eqref{sigmaDot} of the 
      resonant angle at the stable equilibrium and of the 
      aperture \eqref{eqThick} of the resonance.}
    \label{Eq11J2i0}
    \begin{tabular}{ >{$}c<{$} | >{$}c<{$} >{$}c<{$} | >{$}c<{$}  >{$}c<{$}}
      \hline \hline
      C_{20} & a^*_u \text{(km)} & a^*_s \text{(km)} & \text{Period (day)} & \text{Aperture (km)}\\
      \hline
      0 & 544.436 & 556.529 & 1.216\,896 & 134.091\\
      -6.872\,554\,928\times 10^{-2} & 576.353 & 566.066 & 1.254\,289 & 125.726\\
      \hline
      \text{Difference} & 31.917 & 9.537 & 134.614\,h & -8.365
    \end{tabular}
  \end{center}
\end{table}\\
We are now interested in the period of the resonant angle at the stable 
equilibrium ($\sigma^*_s,a^*_s$). This is done by linearizing in a 
neighborhood of the equilibrium. Then the Hamiltonian close to the 
equilibrium is an harmonic oscillator that can be expressed in 
action-angle variables and the frequency at the stable equilibrium 
(the subscript $eq.$ means ``evaluated at the equilibrium'') is given by: 
\begin{equation}\label{sigmaDot}
  \dot{\sigma} = \sqrt{\jacc{\H}{L'}\Big|_{eq.}\,\jacc{\H}{\sigma}\Big|_{eq.}}
  \qquad \text{and}\qquad T=2\pi/\dot{\sigma}\,.
\end{equation}
The values of the period of the resonant angle at the equilibrium 
are gathered in Tab.~\ref{Eq11J2i0}. 
By a similar approach, we can easily estimate the width of the 
resonant zone; we take the Hamiltonian level curve corresponding to one 
of the unstable equilibria $L_u'^*$ and $\sigma_u^*$ noted $\H_u^*$ 
%$\H_u(L_u, \sigma_u,\Lambda)$ 
that we equal to the generic Hamiltonian \eqref{reso11i0} 
giving the dynamics $\sigma$ and $L'$ along that curve: 
%evaluated at $\sigma_s$:
\begin{equation}
%  \H_u(L_u, \sigma_u,\Lambda)
  \H_u^*=-\frac{\mu^2}{2L'^2}- R_e^{2}\,\frac{\mu^4}{L'^6}
  \Bigg\{ -\frac{1}{2}C_{20} + 3 \big(C_{22}\cos(2\sigma)+S_{22}\sin(2\sigma)\big)\Bigg\} - \dot{\theta} L'\,.
\end{equation}
For $\sigma=\sigma^*_s$, we find the maxima ($L'^*_{\max}$) and minima 
($L'^*_{\min}$) of this banana-curve. We obtain the width 
of the banana at the stable points:
\begin{equation}\label{eqThick}
  \Delta L=L'^*_{\max}-L'^*_{\min}\,,
\end{equation}
i.e. the aperture of the resonant zone summarized in Tab.~\ref{Eq11J2i0}.

\subsubsection{Circular polar case: $e=0,\,i=\pi/2$}
The orbit of Dawn is polar and we have seen 
that the inclination does not differ much from $90\Deg$ for all 
the radial distances from the center of Vesta (Section \ref{sectMap}), 
so let us fix $i=\pi/2$.\\
%we only interest to $i=\pi/2$ constant.\\

The methods to find the location of the equilibria, the period at 
the equilibria and the aperture of the resonance are similar to 
that used for the equatorial case (\ref{sec11eqcas}).\\

\noindent In the circular polar case, the Hamiltonian \eqref{HamJ22} becomes 
\begin{equation}\label{reso11i90}
  \H_{1:1}=-\frac{\mu^2}{2L'^2}- R_e^{2}\,\frac{\mu^4}{L'^6}
  \Bigg\{ \frac{1}{4}C_{20} + \frac{3}{4} \big(C_{22}\cos(2\sigma)+S_{22}\sin(2\sigma)\big)\Bigg\} - \dot{\theta}L'\,.  %+ \dot{\theta}(\Lambda'-L')\,.
\end{equation}
The $\sigma$ location of the equlibria is always given by 
Eq~\eqref{sigmaloc11}, i.e. for the particular value of Vesta 
\begin{equation}\label{localisationEq11}
  \sigma^*_{s1}=\pi/2,\quad \sigma^*_{s2}=3\pi/2\quad \text{and} \quad 
  \sigma^*_{s1}=0    ,\quad \sigma^*_{s2}= \pi\,.
\end{equation}
The values of the location, period and aperture are gathered  
in Tab.~\ref{Eq11J2i90}. 
\begin{table}[h!]
  \begin{center}
    \caption{The circular polar case: semi-major axis location 
      of the equilibria (stable and unstable) 
      of the 1:1 resonance (second and third column) 
      with respect to the value of $C_{20}$. 
      In fourth and fifth columns  analytical calculating of the period of the 
      resonant angle at the stable equilibrium and of the aperture of 
      the resonance respectively. We also recall a part of the results of 
      the Tab.~\ref{TabLocalThcikReso} and the Tab.~\ref{TabPriodDgaNum11}. 
      for comparison.}
    \label{Eq11J2i90}
    \begin{tabular}{ >{$}c<{$} | >{$}c<{$} >{$}c<{$} | >{$}c<{$}  >{$}c<{$}}
      \hline \hline
      C_{20} & a^*_u \text{(km)} & a^*_s \text{(km)} & \text{Period (day)} & \text{Aperture (km)}\\
      \hline
      0 & 552.133 & 549.113 & 2.448\,927 & 66.699\\
      -6.872\,554\,928\times 10^{-2} & 540.494 & 537.159 & 2.411\,514 & 69.363\\
%      \text{Difference} & 31.917 & 9.537 & 134.614\,h & -8.365
      \hline
      \text{Difference} & -11.639 & -11.957 & -134.686\,h & 2.664\\
      &&&&\\
      \text{Num. model }&\multicolumn{2}{c|}{\text{Mean of the s.m.a.}}&\text{Period (day)} & \text{Aperture (km)}\\
      \hline
      J_{22} & \multicolumn{2}{c|}{549.63} & 2.45 & 66.6 \\
      J_{22} +C_{20} & \multicolumn{2}{c|}{538.42} & 2.47 & 65.2\\
      \text{Until degree \& order 4} & \multicolumn{2}{c|}{538.89} & 2.56 & 63
    \end{tabular}
  \end{center}
\end{table} 

We bring to the fore that, for an inclination of $90\Deg$, the period 
at the stable equilibria and the aperture (Tab.~\ref{Eq11J2i90}) 
of the resonance are respectively equal to the double and the half 
%of the results found for the equatorial ($i=0\Deg$) case 
of the equatorial results 
(Tab.~\ref{Eq11J2i0}). The doubling of the period can be easily 
understood regarding the equations \eqref{reso11i0}, \eqref{reso11i90} 
and \eqref{sigmaDot}: in the Hamiltonian \eqref{reso11i90} 
a $\frac{1}{4}$ factor appears, just before the $C_{22}$ term, 
(that does not exist in the Hamiltonian \eqref{reso11i0}) that 
becomes $\frac{1}{2}$ after the square root of the frequency 
equation \eqref{sigmaDot}, which gives a factor $2$ for the period. 
The aperture can be explained in a similar way. 

We also point out that the location of the resonance is shifted of $\sim 7$~km 
from the equatorial case for $C_{20}=0$ and of $\sim 30$~km for $C_{20}\neq 0$. 
If we neglect the $C_{20}$ term, we introduce an error of about $2.6$~km 
on the aperture of the resonance and of about $11.8$~km on the 
location of the equilibria. This last result coincides well 
with the $11.2$~km numerically found in Tab.\ref{TabPriodDgaNum11}.

A comparison of the analytical results with the numerical ones 
shows a good agreement 
(Tab.~\ref{TabLocalThcikReso}, \ref{TabPriodDgaNum11} and \ref{Eq11J2i90}). 
%If we compare these analytical results with the numerical results 
%we can notice that the analytical results are quite good. 
The difference between the numerical and the analytical results comes 
from the approximation of the osculating elements by the mean keplerian ones. 
%difference between the osculating and the mean keplerian elements. 
%The aperture results from short period contributions that are actually 
%averaged in the analytical study. 
%For the aperture the difference can be come from the more complet model 
%used in the numerical integration.
The different apertures of the resonances should also come from 
contributions that have been neglected or averaged in the analytical study.

%the more 
%complet model used in the numerical integration.

\subsubsection{Addition of $C_{40}$}
\begin{table}[htb]
  \begin{center}
    \caption{For the equatorial (noted Eq.) and the polar (noted Pol.) cases: 
      semi-major axis location of the stable equilibria of the 
      1:1 resonance, period of the resonant angle at the stable equilibrium 
      and aperture of the resonance with respect to the terms used 
      in the Hamiltonian.}
    \label{Eq11J2J4i90}
    \begin{tabular}{lr|>{$}c<{$} >{$}c<{$} >{$}c<{$}}
        \hline \hline
        Term &  & a^*_s \text{(km)} & \text{Period (day)} & \text{Aperture (km)}\\
        \hline
%        \multirow{2}{*}{$C_{22}$} & Eq. & 544.436 & 1.216\,896 & 134.091\\
%                                 & Pol.& 549.113 & 2.448\,927 &  66.699\\
        \multirow{2}{*}{$C_{22}+C_{20}$} & Eq. & 566.066 & 1.254\,289 & 125.726\\
                                       & Pol.& 537.159 & 2.411\,514 &  69.363\\
        \multirow{2}{*}{$C_{22}+C_{20}+C_{40}$} & Eq. & 567.035 & 1.253\,964 & 125.175\\
                                             & Pol.& 537.671 & 2.410\,739 &  69.177
      \end{tabular}
  \end{center}
\end{table}
We have seen that the addition of the $C_{20}$ term has a strong 
influence on the location of the resonance. Then, here, we test 
the effect of the $C_{40}$ term: 
\begin{equation}\label{J4term}
  \H_{J_{44}}=\H_{1:1}-R_e^4 \frac{\mu^6}{L'^{10}}\left(\frac{105}{64}\sin^4 i-\frac{15}{8}\sin^2 i+\frac{3}{8}\right)\left(1+\frac{3}{2}e^2\right)C_{40}\,.
\end{equation}
We summarize the values of the location, period and aperture in 
Tab.~\ref{Eq11J2J4i90}. We can see that this term does not significantly 
affect the results.

\subsubsection{Effect of the error on the coefficient $C_{22}$ and $C_{20}$}
It can be shown \citep{Balmino1994} that
\begin{equation}
  - J_2 = C_{20} =\frac{1}{5R_e^2}\left(c^2-\frac{a^2+b^2}{2}\right) \quad 
  \text{ and } \quad C_{22} =\frac{1}{20R_e^2}(a^2-b^2)\,,
\end{equation}
where $a,b$ and $c$ (with $a \leq b \leq c$) are the semi-axes of 
the asteroid and $R_e$ is the mean equatorial asteroid radius. 
If we take the values of $a,b,c,R_e$ of \citet{Tricarico2010}: 
\begin{equation}
  a=289\pm5\,\text{km}, \qquad   b=280\pm5\,\text{km}, \qquad 
  a=229\pm5\,\text{km}, \qquad  R_e=300\,\text{km}\,,
\end{equation}
then the error on the value of $C_{20}$ and $C_{22}$ is 
\begin{equation}\label{ErrC22}
  \frac{\Delta C_{20}}{C_{20}}=18 \%=\frac{\Delta \bar{C}_{20}}{\bar{C}_{20}}
  \qquad \text{and} \qquad 
  \frac{\Delta C_{22}}{C_{22}}=111\%=\frac{\Delta \bar{C}_{22}}{\bar{C}_{22}} \,.
\end{equation}
If we slightly change the reference value $R_e$ or if we add 
an error (e.g. $\pm 5$~km) on this reference radius, 
the results \eqref{ErrC22} are similar. 
The second value of \eqref{ErrC22} can be surprising but it gives 
an idea of the boundary on the values of $C_{22}$. Owing to 
the supposed constant density of the asteroid, the $C_{22}$ coefficient is 
assumed to be positive. Then the minimal (non null) value of $C_{22}$ 
is arbitrarly chosen equal to $11\%$ of the initial value of $C_{22}$. 
%we arbitrarly take $11\%$ of $C_{22}$ as minimal 
%value.
So, we calculate the extremal values and for each of them we 
analytically evaluate the position (semi-major axis) of the 1:1 
resonance \eqref{EqEquilL7}, the period at the equilibrium 
\eqref{sigmaDot} and the aperture of the resonance \eqref{eqThick} 
for the polar circular case. The results are presented 
in Tab.~\ref{TabC20C22Val}. The location in semi-major axis 
is between $532$ and $541$~km. 
On one hand, for a chosen value of $\bar{C}_{22}$, 
the maximum errors on the evaluation of the position of the 
resonance, of the period and of the aperture are respectively equal to 
$4.6$~km, $0.04$ d and 1~km, due to the error on the $\bar{C}_{20}$. 
On the other hand, for a fixed value of $\bar{C}_{20}$, the extremal 
values of $\bar{C}_{22}$ produce errors equal 
to $3.5$~km, $5.6$~days and $78$~km. So it is very important to increase 
the accuracy on the estimation of the three semi-axes ($a,b,c$). Now if 
we suppose this improvement, for example an error of $\pm 1$~km on 
$a,b$ and $c$, the error on $\bar{C}_{20}$ and $\bar{C}_{22}$ becomes 
respectively $4\%$ and $22\%$. Then the extremal values for the 
location, period and aperture are $[536.28 - 538.031]$~km, 
$[2.181 - 2.733]$~days and $[61.128 - 76.780]$~km.

\begin{table}[htb]
  \begin{center}
    \caption{The circular polar case: analytical values of the 
      location (semi-major axis noted s.m.a.) of the stable equilibrium of 
      the 1:1 resonance (s. s.m.a), of the period at the equilibrium (T) and 
      of the aperture (Apert.) of the resonance with respect to 
      the extremal values of the $\bar{C}_{20}$ and $\bar{C}_{22}$. 
      The middle column and the middle line (T\&S) are calculated 
      with the values of $\bar{C}_{20}$ and $\bar{C}_{22}$ used in this 
      paper and found in \citet{Tricarico2010}. 
      We emphasized the T\&S (blue), maximum (red) and 
      minimum (green) values obtained for the semi-major axis 
      location, the period and the aperture. \label{TabC20C22Val}}
    \begin{tabular}{l | l | >{$}c<{$} >{$}c<{$}  >{$}c<{$} | l }
      \multicolumn{2}{c|}{} & \text{Min.} & \text{T\&S} & \text{Max.}\\
      \cline{2-5}
       & \backslashbox{$\bar{C}_{22}$}{$\bar{C}_{20}$} & -0.025\,203 & -0.030\,735 & -0.036\,267 &\\
      \hline
      \multirow{3}{*}{Min.} & \multirow{3}{*}{$5.248\times10^{-4}$} & \red{540.870}   & 538.655 & 536.397  & s. s.m.a. (km)\\
      &                                                            & \red{7.305}     & 7.284   & 7.264    & T (d)\\
      &                                                            & \green{22.793}  & 22.964  & 23.142   & Apert. (km) \\
      \hline
      \multirow{3}{*}{T\&S} & \multirow{3}{*}{$0.004\,771$} & 539.402 & \blue{537.159} & 534.871 & s. s.m.a. (km) \\
      &                                                           & 2.418   & \blue{2.412}   & 2.405   & T (d) \\
      &                                                           & 68.841  & \blue{69.363}  & 69.906  & Apert. (km) \\
      \hline
      \multirow{3}{*}{Max.} & \multirow{3}{*}{$0.010\,067$} & 537.545 & 535.265 & \green{532.938} & s. s.m.a. (km)\\
      &                                                     & 1.661   & 1.656   & \green{1.652}   & T (d) \\
      &                                                     & 100.221 & 100.991 & \red{101.194}   & Apert. (km) 
    \end{tabular}
  \end{center}
\end{table}

\subsection{Resonance 1:2, 2:3 and 3:2 in the polar case}

For the 1:2, 2:3 and 3:2 resonance cases (in the circular polar case), we 
proceed in the same way as in the 1:1 resonance case to find the 
Hamiltonian, the location of the equilibria, the periods at the stable 
equilibria and the aperture of the resonances.\\

First, for each of the resonances (1:2, 2:3 and 3:2), 
we determine the values of the integers $n,m,p,q$ to obtain the 
resonance $q_1:q_2$ to get Kaula gravitational argument including the 
resonant angle $\sigma=q_2\lambda - q_1 \theta$. 
% but not explicitly including the mean anomaly $M$. 
%Let us that this last condition is equivalent to keep all 
%solution $n,m,p,q$ such as $\sigma=q_2\lambda - q_1 \theta$, 
%after which we average over the mean anomaly $M$. 
%After this averaging, the remaining terms are those 
%that dit not explicitly contains the mean anomaly from the start. 
In Tab.~\ref{TabResoAngle}, we gather 
the resonances, the combinations $n,m,p,q$, the associated 
resonant angles, the Kaula gravitational arguments and 
the functions $F_{nmp}(i)$ and 
$G_{npq}(e)$ respectively evaluated at $i=\pi/2$ and $e=0$.
\begin{table}[ht]
  \begin{center}
    \caption{Expression of the resonant angles $\sigma$, the Kaula 
    gravitational arguments $\Theta_{n,m,p,q}$, the functions $F_{nmp}(i=\pi/2)$ 
    and $G_{npq}(e=0)$ with respect to the values of $n,m,q=0$ for 
    the resonant case $q_1:q_2$. \label{TabResoAngle}}
    \begin{tabular}{c c c c c | >{$}c<{$} >{$}c<{$} | >{$}c<{$}  >{$}c<{$} }
      \hline \hline
      $q_1:q_2$ &n & m & p & q & \sigma & \Theta_{nmpq} & F_{nmp}(i=\pi/2) & G_{npq}(e=0)\\
      \hline
      1:2 & 4 & 1 & 1 & 0 & 2(\Omega+\omega+M)-\;\,\theta & \sigma-\Omega & 5/16  & 1\\
      2:3 & 3 & 2 & 0 & 0 & 3(\Omega+\omega+M)-2\theta & \sigma-\Omega & 15/8  & 1\\
      3:2 & 4 & 3 & 1 & 0 & 2(\Omega+\omega+M)-3\theta & \sigma+\Omega & 105/8 & 1
    \end{tabular}
  \end{center}
\end{table}

Second, thanks to this combination of $n,m,p,q$, we determine 
%a symplectic transformation % equivalent to \eqref{symplectictransfo}. Below 
%we present 
the different sets of canonical variables, 
equivalent to Eq.~\eqref{canonicalvariable}, for each resonance:
\begin{eqnarray}
  \text{Resonance 1:2} & \quad \quad \sigma=2\lambda-\;\theta, \quad & L=2L', \quad \theta' = \theta, \quad \Lambda=\Lambda'-L'\\
  \text{Resonance 2:3} & \quad \quad \sigma=3\lambda-2\theta, \quad & L=3L', \quad \theta' = \theta, \quad \Lambda=\Lambda'-2L'\label{canonicalvariableRes23}\\
  \text{Resonance 3:2} & \quad \quad \sigma=2\lambda-3\theta, \quad & L=2L', \quad \theta' = \theta, \quad \Lambda=\Lambda'-3L'\,.
\end{eqnarray}
Then we can write the Hamiltonians assiocated to these resonances:
\begin{eqnarray}
  \H_{1:2}&=&-\frac{\mu^2}{8L'^2} - \frac{\mu^4\,R_e^2\,C_{20}}{256\,L'^6} \\
  &&-\frac{5}{16384}\frac{\mu^6 R_e^4}{L'^{10}} 
  \;\;\Big(-S_{41}\cos(\sigma-\Omega)+C_{41}\sin(\sigma-\Omega)\Big) + 
  \dot{\theta}(\Lambda'-L')\nonumber\\
  \H_{2:3}&=&-\frac{\mu^2}{18L'^2} - \frac{\mu^4\,R_e^2\,C_{20}}{2916\,L'^6}\label{Ham23}\\
  &&-\frac{5}{17496}\frac{\mu^5 R_e^3}{L'^{8}} 
  \;\;\Big(-S_{32}\cos(\sigma-\Omega)+C_{32}\sin(\sigma-\Omega)\Big) + 
  \dot{\theta}(\Lambda'-2L')\nonumber\\
  \H_{3:2}&=&-\frac{\mu^2}{8L'^2} - \frac{\mu^4\,R_e^2\,C_{20}}{256\,L'^6} \\
  &&-\frac{105}{8192}\frac{\mu^5 R_e^4}{L'^{10}} 
  \;\;\;\,\Big(-S_{43}\cos(\sigma+\Omega)+C_{43}\sin(\sigma+\Omega)\Big) + 
  \dot{\theta}(\Lambda'-3L')\nonumber\,.
\end{eqnarray}
In comparison to the Hamiltonian \eqref{reso11i90}, the term in front of 
the cosine is $-S_{nm}$ instead of $C_{nm}$. That is due to the odd value 
of the $n-m$ ($4-1=3$, $3-2=1$, $4-3=1$) taking place in 
equation \eqref{Snmpq}.

With the same procedure as in the 1:1 resonance, we can determine the 
position of the stable equilibria in the $\sigma$ variable:
\begin{eqnarray}
  \text{Reso. 1:2 }\quad & \sigma_s = &\;\;\,\Omega-\arctan\left(\frac{C_{41}}{S_{41}}\right)+(2k+1)\pi\\% = \;\;\, \Omega\Deg+ 78.102\,773\Deg  \\
  \text{Reso. 2:3 }\quad & \sigma_s = &\;\;\,\Omega-\arctan\left(\frac{C_{32}}{S_{32}}\right)+2k\pi\\% \qquad\;\,= \;\;\; \Omega\Deg+ 79.357\,931\Deg \\
  \text{Reso. 3:2 }\quad & \sigma_s = &-\Omega-\arctan\left(\frac{C_{43}}{S_{43}}\right)+(2k+1)\pi% = -\Omega\Deg+ 94.936\,520\Deg \,,
\end{eqnarray}
and the unstable equilibria:
\begin{eqnarray}
  \text{Reso. 1:2 }\quad &  \sigma_u = &\;\;\,\Omega-\arctan\left(\frac{C_{41}}{S_{41}}\right)+2k\pi\\% \qquad\;\,= \;\;\, \Omega\Deg -258.102\,773\Deg\\
  \text{Reso. 2:3 }\quad &  \sigma_u = &\;\;\,\Omega-\arctan\left(\frac{C_{32}}{S_{32}}\right)+2(k+1)\pi\\% = \;\;\, \Omega\Deg -259.357\,931\Deg\\
  \text{Reso. 3:2 }\quad &  \sigma_u = &-\Omega-\arctan\left(\frac{C_{43}}{S_{43}}\right)+2k\pi\,.% \qquad\;\,= -\Omega\Deg -85.063\,479\Deg\,.
\end{eqnarray}
For clarity, we give the mean longitudes corresponding to the 
stable equilibria ($\lambda_s$), in degrees: 
\begin{eqnarray}
  \text{Reso. 1:2}\quad & \lambda_s = & 129.05\Deg+\frac{\theta+\Omega}{2}+k\,180\Deg, \,\quad k\in\{0,1\}\\ %\;\;\,\frac{1}{2}\Omega-\frac{1}{2}\arctan\left(\frac{C_{41}}{S_{41}}\right)+\frac{2k+1}{2}\pi+\frac{1}{2}\theta = \\
  \text{Reso. 2:3}\quad & \lambda_s = & 26.45\Deg+\frac{2\theta+\Omega}{3}+k\,120\Deg, \quad k\in\{0,1,2\}\label{EquiRes23} \\%\;\;\,\frac{1}{3}\Omega-\frac{1}{3}\arctan\left(\frac{C_{32}}{S_{32}}\right)+\frac{2}{3}k\pi+\frac{2}{3}\theta\\
  \text{Reso. 3:2}\quad & \lambda_s = & 47.47\Deg+\frac{3\theta-\Omega}{2}+k\,180\Deg, \quad k\in\{0,1\}% -\frac{1}{2}\Omega-\frac{1}{2}\arctan\left(\frac{C_{43}}{S_{43}}\right)+\frac{2k+1}{2}\pi+\frac{3}{2}\theta\,;
\end{eqnarray}

The values of the locations, periods and apertures of the resonances are 
summarized in Tab.~\ref{EqOthResoi90}.
\begin{table}[htb]
  \begin{center}
    \caption{The circular polar case: resonances 1:2, 2:3 and 3:2. 
      Values of the semi-major axis location 
      (stable and unstable equilibria), the period of the 
      resonant angle at the stable equilibrium and the aperture of 
      the resonance with respect to the terms used in the Hamiltonian. 
%      for each 
      For comparison, the results of the numerical test of 
      the Tab.~\ref{TabLocalThcikReso} (noted by Num. in the first column) 
      are recalled.}
    \label{EqOthResoi90}
    \begin{tabular}{ l >{$}c<{$}| >{$}c<{$} >{$}c<{$} | >{$}c<{$}  >{$}c<{$}}
      \hline \hline
      Reso.& & a^*_u \text{(km)} & a^*_s \text{(km)} & \text{Period (day)} & \text{Aperture (km)}\\
      \hline
      \multirow{3}{*}{1:2}& C_{20}=0 & 874.090 & 874.057 & 76.341\,231 &  6.796\\
      & C_{20}\neq 0 & 866.882 & 866.847 & 75.251\,862 & 9.953\\
      & \text{Num. } & \multicolumn{2}{c|}{871.5} & -- & 7.5\vspace{0.1cm}\\
      \multirow{3}{*}{2:3}& C_{20}=0 & 721.698 & 721.365 &  9.755\,028 & 21.953\\
      & C_{20}\neq 0 & 712.898 & 712.541 & 9.608\,888 & 22.575\\
      & \text{Num. } & \multicolumn{2}{c|}{717.3} & -- & 22.5\vspace{0.1cm}\\
      \multirow{3}{*}{3:2}& C_{20}=0 & 420.508 & 419.911 &  4.147\,596 & 20.056\\
      & C_{20}\neq 0 & 404.528 & 403.755 & 3.885\,027 & 22.401\\
      & \text{Num. } & \multicolumn{2}{c|}{409.1} & -- & 18 
    \end{tabular}
  \end{center}
\end{table}

We notice that, if we neglect the $C_{20}$ terms, we introduce a mean error of 
$\sim 11$~km on the location of the resonances, of $\sim  0.5$~days 
on the periods at the stable equilibria and of $\sim  2$~km on the 
apertures of the resonances. If we compare (location and aperture) 
these analytical results with the numerical results 
(Tab.~\ref{TabLocalThcikReso} repeated in the Tab.~\ref{EqOthResoi90}), 
we point out that the analytical results are adequate.

%\section{Increase of the eccentricity in the resonance 1:1 and 2:3}\label{secEgd}
\section{Increase of the eccentricity in the 2:3 resonance}\label{secEgd}
In section~\ref{sectAnalyt}, we have analytically found the 
positions of the main resonances, their apertures and their periods at 
the stable equilibria. As shown in Fig.~\ref{C22PhaseSpace}, the 
ground-track resonances show an oscillation of the semi-major axis 
and of the resonant angle. This is why, in the left panel of 
Fig.~\ref{DgaExcRayDawn}, we see an increase of the variation of the 
semi-major axis exactly at the initial semi-major axis, corresponding 
to the ground-track resonances. 
%But we did not explained the increase of 
%the eccentricity that we see at the center of the resonance 1:1 and 
%at the boundary of the resonance 2:3 (central panel of 
%Fig.~\ref{DgaExcRayDawn}). 
%% In this section we investigate this phenomenon. 
%% We separately study the resonance 1:1 and 2:3. For each we make some 
%% numerical test (we present here only the conclusive results) 
%% to understand what affects the increase of the eccentricity. 
%% After what we give the corresponding Hamiltonian. 
Up to now we have not explained that the strongest perturbations on 
the radial distance come from the 2:3 resonance 
(right panel of Fig.~\ref{DgaExcRayDawn}). 
In this section, we investigate this phenomenon.

\begin{figure}[htbp] 
  \begin{center}
    \includegraphics[draft=false,width=\textwidth]{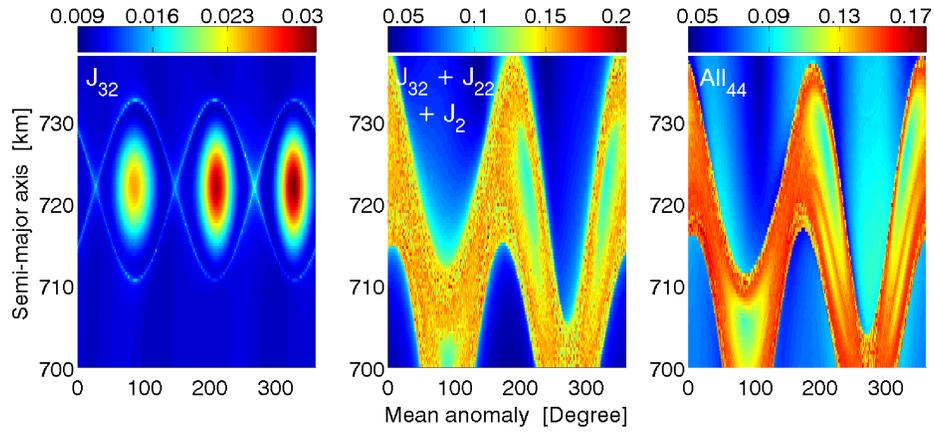}
    \caption{\label{FigRes23E} Around the 2:3 ground-track resonance. 
      The amplitude of variation of the eccentricity computed as a 
      function of the initial mean anomaly and the initial 
      semi-major axis. The anomaly step is $1\Deg$ and the 
      semi-major axis step is $400$~m. The initial conditions are 
      $i_0 = 90\Deg$, $\Omega=\omega=0\Deg$ and $\theta=0\Deg$.
      The integration time is 1 year. 
      %from epoch fixed at 14 September 2019. 
      The equations of motion include the central body attraction, 
      the harmonic $J_{32}$ in the left panel; 
      the harmonics $J_{22}$ and  $J_{32}$ in the central panel; 
      and all of the harmonics untill degree and order 4 ($\text{All}_{44}$) 
      in the right panel.}
  \end{center}
\end{figure} 

We make a huge number of numerical tests but we present here only 
the significant ones. In Fig.~\ref{FigRes23E}, we report the results 
of a numerical integration for a set of $34\,920$ orbits, propagated over a 
time span of 1~year. We consider a set of initial 
conditions defined by a mean anomaly grid of $1\Deg$ and a
semi-major axis grid of $400$~m, spanning the $[700, 738.8]$~km range 
around the 2:3 resonance. 
The other fixed initial conditions are $i_0 = 90\Deg$ for the inclination,
$\Omega=\omega=0\Deg$ for the longitude of the ascending node and the 
argument of pericenter, and $\theta=0\Deg$ for the sideral time. 
%at epoch 14 September 2019.
The color code represents the maximal value of the eccentricty variation 
($e_{\max}-e_{\min}$) during the integration.

In the left panel of Fig.~\ref{FigRes23E}, we only use the $J_{32}$ 
term to modelize the central body. As expected by Eq. \eqref{EquiRes23},
 we see three pendulum-like phase spaces. 
In the central panel, we add the $J_2$ and the $J_{22}$ terms. We notice 
that the eccentricity reaches higher values. In the right panel, 
the gravity field contains all of the terms until degree and  order~4. 
The results are similar to these of the central panel. 
Then, with the help of Fig.~\ref{DgaExcRayDawn} and Fig.~\ref{FigRes23E}, 
we can conclude that the high value of the amplitude of variation of 
the radial distance that we see in Fig.~\ref{DgaExcRayDawn} is due to 
an increase of the eccentricity in the 2:3 resonance. 
We can also deduce from Fig.~\ref{FigRes23E} that the major part of this 
increase of the eccentricity is due to the $J_{22}$ coefficient. 

%As expected with the $J_{32}$ term, we see three \eqref{EquiRes23}
%pendulum-like phase spaces (left panel of Fig.~\ref{FigRes23E}). 
%%With only the $J_{22}$ term we see an increase of the eccentricity 
%%(right upper panel) just 
%%under the semi-major axis corresponding to the resonance 2:3. 
%Then if we combine the $J_{32}$ and the $J_{22}+J_2$ terms we obtain the three 
%eyes of the resonance 2:3 with an increase of the eccentricity 
%around the separatrix (central panel of Fig.~\ref{FigRes23E}). 
%% Afterwards we add the other terms until order and degree 4 (right panel).  
%% A comparison between the central and the right panel shows a good agreement. 
%% We make some other numerical tests but we present here only the 
%% significant ones 
%% and we conclude that the major part of the increase 
%% of the eccentricity around the separatrix of the 2:3 resonance is due 
%% to the $J_{22}$ coefficient. 

Now we analytically justify the fact that the first coefficient of the 
gravity field involving the eccentricity is the $J_{22}$ term. 
First, we determine the values of the integers $n, m, p, q$ to obtain 
the 2:3 resonance. Second, we only keep the non-null Kaula gravitational 
argument including the resonant angle $\sigma=3\lambda-2\theta$. 
% but not explicitly including the mean anomaly $M$. 
In Tab.~\ref{TabRes23TFG} we present the combinations $n, m, p, q$, 
the associated Kaula gravitational argument and the functions $F_{nmp}(i)$ 
(evaluated at $i = \pi/2$) and $G_{npq}(e)$.

\begin{table}[htb]
  \begin{center}
    \caption{Expression of the functions $\Theta_{nmpq}\neq 0, F_{nmp}(i=\pi/2)$ and
      $G_{npq}(e)$ with respect to the values of $n, m, p, q$ for the 
      resonance 2:3. }
    \label{TabRes23TFG}
    \begin{tabular}{r r r r | >{$}l<{$} | >{$}r<{$} | >{$}r<{$} }
      \hline \hline
      n & m & p & q & \qquad\Theta_{nmpq} & F_{nmp}(i=\pi/2) & G_{npq}(e)\qquad\qquad\qquad\\
      \hline
      2 & 2 & 0 &  1 & 3\lambda-2\theta-\omega-\Omega&  3/4   \quad\qquad& 7e/2-123e^3/16+\dots\\
      3 & 2 & 0 &  0 & 3\lambda-2\theta-\Omega       &  15/8  \quad\qquad& 1-6e^2+423e^4/64+\dots\\
      4 & 2 & 0 & -1 & 3\lambda-2\theta+\omega-\Omega& -105/32\quad\qquad& -3e/2+75e^3/16+\dots\\
      4 & 2 & 1 &  1 & 3\lambda-2\theta-\omega-\Omega& -15/8  \quad\qquad& 9e/2-3e^3/16+\dots
    \end{tabular}
  \end{center}
\end{table}

The first line contains the first function $G_{npq}$ directly proportional 
to $e$, it is related to the $J_{22}$ coefficient. The second line 
corresponds to the coefficient $J_{32}$ that induces the 2:3 resonance. 
In this term, the eccentricity takes place at the second order $e^2$. 
The last two lines are also directly proportional to $e$ but, for Vesta, 
the coefficient $C_{44}$ is 30 times smaller than the coefficient $C_{22}$.

If we add the coefficients $C_{22}$ and $S_{22}$ to 
the Hamiltonian~\eqref{Ham23} and if 
we keep only the zeroth and first orders in eccentricity, we obtain:
\begin{eqnarray}
  \H_{2:3}&=&-\frac{\mu^2}{2L^2} 
  -\frac{15\mu^5 R_e^3}{8L^8}\Big(-S_{32}\cos(\sigma-\Omega)+C_{32}\sin(\sigma-\Omega)\Big)\\
  &-& e \frac{21\mu^4 R_e^2}{8L^6}\Big(C_{22}\cos(\sigma-\omega-\Omega) + S_{22}\sin(\sigma-\omega-\Omega)\Big)  +\dot{\theta}\Lambda\,. \nonumber
\end{eqnarray}

With this Hamiltonian we can qualitatively give the secular 
equation of the eccentricity $\dot{e}\approx |\dpar{\H}{\omega}|$. 
To have an idea of the variation of the eccentricity during the motion 
we can look at the $\dot{e}/e$ value:
\begin{equation}
  \frac{\dot{e}}{e}\approx \left|\dpar{\H}{\omega}\right|\;\frac{1}{e} 
  = \frac{21\mu^4R_e^2}{8L^6} 
  \Big( C_{22}\sin(\sigma-\omega-\Omega)- S_{22}\cos(\sigma-\omega-\Omega)\Big)\,.
\end{equation}
Then the variation of the eccentricity depends on one of the biggest 
coefficients of the gravity field $C_{22}$. For Vesta, this 
coefficient $C_{22}$ is 5 times greater than the coefficient $C_{32}$ 
responsible for the 3:2 ground-track resonance. 
We evaluate (for the values linked to Vesta), near to the 3:2 resonance, 
the multiplying coefficient appearing in front of the 
two major contributions $C_{32}$ and $C_{22}$:
\begin{equation}
  \frac{15\mu^5 R_e^3}{8L^8}C_{32}\approx 6913 \qquad \text{and} \qquad 
  e \frac{21\mu^4 R_e^2}{8L^6}C_{22}\approx e\;25179\,.
\end{equation}
The second value is $3.6$ times greater than the first. 
In the case of Dawn orbiting Vesta, this highlights the importance 
of this additional term  $C_{22}$ in the 2:3 resonance. 

\section{Thrust and probability of capture}\label{secThrust}

Due to the low thrust propulsion, Dawn will slowly cross the 
different ground-track resonances previously detected. During this transit, 
the orbit could become chaotic, or Dawn can be trapped in a resonance. 

\begin{figure}[htbp] 
  \begin{center}
    \includegraphics[draft=false,width=\textwidth]{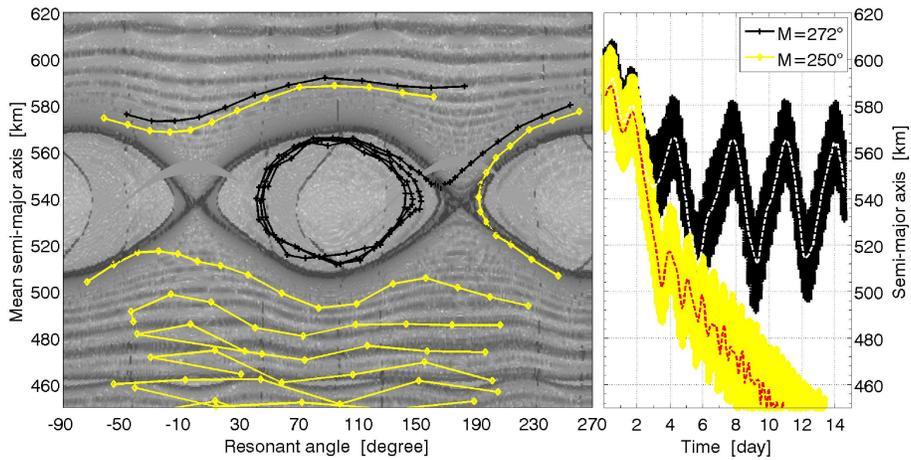}
    \caption{\label{ThrustCapture} Evolution of Dawn from HAMO to LAMO with 
      a thrust of $-20$mN with two different initial mean anomalies. 
      In the right and left panels, the dashed lines (respectively the 
      continued) represent the evolution of the mean semi-major axis 
      with respect to the time (respectively to the resonant angle). 
      In the right panel, we overlay the evolution 
      of the osculating semi-major axis (continued lines). 
      The trapping in the resonance 
      1:1 depends on the phase of the resonant angle at the time of 
      crossing the separatrix. 
      The force model contains the gravity field until degree and order 4 
      and a thrust of $20$mN opposite to the Vesta rotation.}
  \end{center}
\end{figure}

\begin{figure}[htbp] 
  \begin{center}
    \includegraphics[draft=false,bb=161 291 452 558,width=0.45\textwidth]{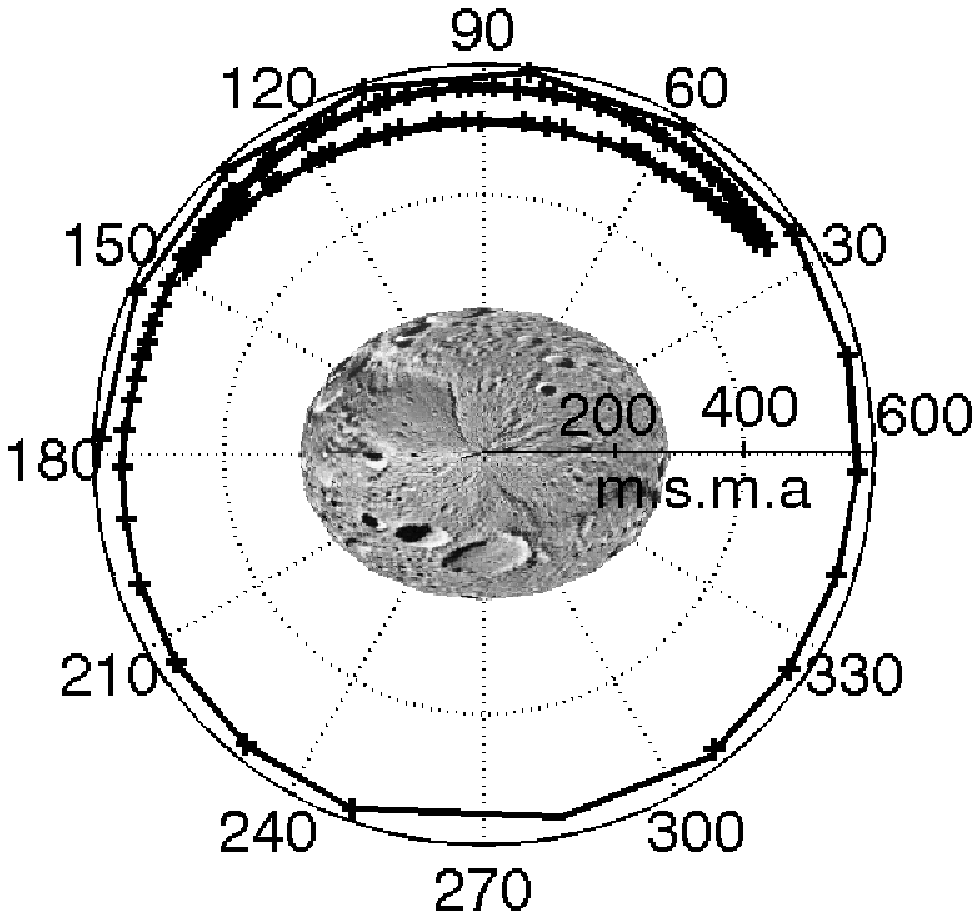}
    \caption{\label{FigCapturePol} 
      In a Vesta centered rotating reference frame as seen from the pole,
      capture in the 1:1 resonance corresponding to the dark line 
      in Fig.~\ref{ThrustCapture}. The radius and the angle are respectively 
      in km and degree.}
  \end{center}
\end{figure}

In this section, we first reproduce the results of \citet{Tricarico2010} 
that we explore more in depth. We simulate the slow descent of Dawn from HAMO, 
using a thrust of 20mN, to LAMO. An initial polar circular orbit and a 
continuous thrusting are assumed, with the direction of the thrust 
opposite to that of the relative velocity of Vesta. 
%The results of the simulation are shown in Fig.~\ref{ThrustCapture}. 
The simulations are displayed in Fig.~\ref{ThrustCapture}. 
In the right panel we draw  
the evolution (continued lines) of the (osculating) semi-major axis 
with respect to the time. The dark and light continued curves 
correspond to an initial mean anomaly respectively equal to $272\Deg$ and 
$250\Deg$. On the same panel we plot with dashed lines the corresponding 
evolution of the {\bf mean} semi-major axis. 
In the left panel, the gray background represents the eyes of the 
resonance 1:1. To clarify, we plot the evolution of the mean semi-major 
axis in the phase space ($\sigma,a$). We notice that the capture in 
resonance depends on the initial mean anomaly. Actually the probability 
of capture depends on the phase of the resonant angle at the time of the 
separatrix crossing. The trapping occurs when the spacecraft crosses 
the 1:1 resonance near the longitude where the Vesta equatorial 
semi-axis is the shortest. The reason is that this is 
for this value ($\sigma=90\Deg$) 
that we analytically find the stable point of the 1:1 resonance 
(Eq.~\ref{localisationEq11}). In Fig.~\ref{FigCapturePol}, we 
plot the orbit (dark line), which leads to a capture in the 1:1 resonance, 
in a polar coordinate projection. We rediscover the banana curve 
of the %illustration of the 1:1 resonant in 
Fig.~\ref{C22PhaseSpace}. 

\begin{figure}[htbp] 
  \begin{center}
    \includegraphics[draft=false,width=\textwidth]{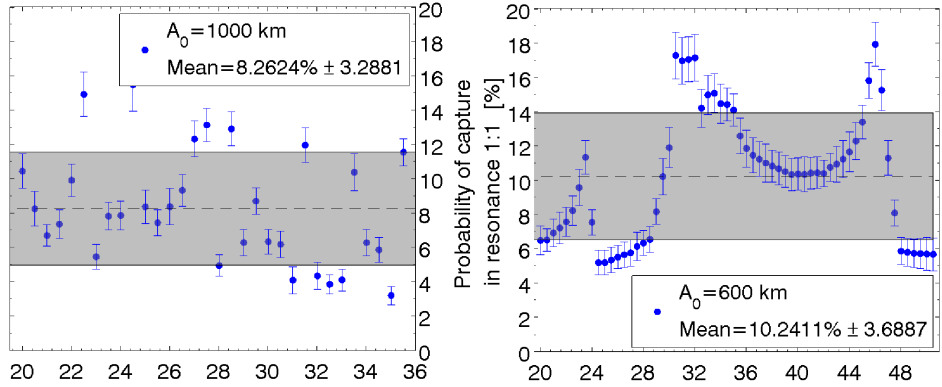}
    \caption{\label{Thrust1000et600} Probability of trapping of Dawn in 
      the 1:1 resonance with respect to the thrust and the initial 
      semi-major axis. For each value of the thrust we make 
      numerical integrations with $1\,000$ different mean anomalies 
      chosen randomly. We repeat this operation 50 times to 
      obtain the means and the standard deviations.}
  \end{center}
\end{figure}

\citet{Tricarico2010} found, with twelve tests, that the probability of 
capture in the 1:1 resonance is near to $1/12\approx 8.33\%$ and they 
observed that higher thrusting than 50mN could exhibit a similar behavior. 
They also noted that the amplitude of the radial oscillations 
increases when crossing the 2:3 resonance, pumping up the eccentricity 
of the orbit. The right panel of our Fig.~\ref{DgaExcRayDawn} also 
presents these structures and we have explained the origin of this phenomenon 
%This last remark is explained, in our Fig.~\ref{DgaExcRayDawn}, 
%by the high amplitude of variation of the eccentricity that we have seen 
%at the 2:3 resonance (Section~\ref{secEgd}). 
in the previous section. 
So it is possible that the increase of the 
eccentricity affects the probability of capture. 
%So we interest to know better 
Let us calculate the probability of capture with respect to the thrust 
and the initial semi-major axis. The results are presented in the 
Fig.~\ref{Thrust1000et600}. To obtain these figures, we take a force model 
that contains the gravitiy field until degree and order 4 and a thrust. 
The thrust is spread from 20mN to 36mN (to 50 for the right panel) with 
a step of 0.5mN. For each value of the thrust, we numerically integrate 
50 times $1\,000$ orbits with a random initial mean anomaly. In the left 
panel, the initial semi-major axis is equal to $1\,000$ km corresponding 
to the HAMO altitude. Then these orbits cross the 2:3 resonance. In the right 
panel the initial semi-major axis is equal to $600$ km and does not cross any 
resonance except the 1:1.

In the left panel, we notice that the probability of capture 
is not linked to the value of the thrust. The mean probability 
of capture is equal to $8.26\%\,\pm3.288$. This confirms the results 
of \citet{Tricarico2010}. In the right panel, the probability of capture 
fluctuates between $4\%$ and $18\%$ with a mean equal to $10.24\%$ 
and a standard deviation of $3.688$. We observe that the probability 
of trapping is directly linked to the value of the thrust. From a thrust 
equal to $48$mN, the probability decreases. So, for any 
initial semi-major axis and mean anomaly, we advice to take 
a thrust between $20$ and $31$mN or greater than 48mN. In this way, 
we obtain an higher probability to cross, without 
permanent trapping, the 1:1 resonance.\\

Now, if the probe is trapped in the 1:1 resonance, it is interesting 
to know how we can get the probe out of the resonance. First, 
in Fig.~\ref{ThrustCapture}, for the captured orbit (dark line) 
in the eye of the resonance, we carry on the thrust of $20$mN. Thus, 
to escape from this resonance, we increase the value of the thrust. 
We decide to increase the thrust from 20 to 35mN 
(as \citealt{Tricarico2010}). We perform two tests where we modify the 
thrust at two different times: after $12$ and $13$ days. 
\begin{figure}[htbp] 
  \begin{center}
    \includegraphics[draft=false,width=\textwidth]{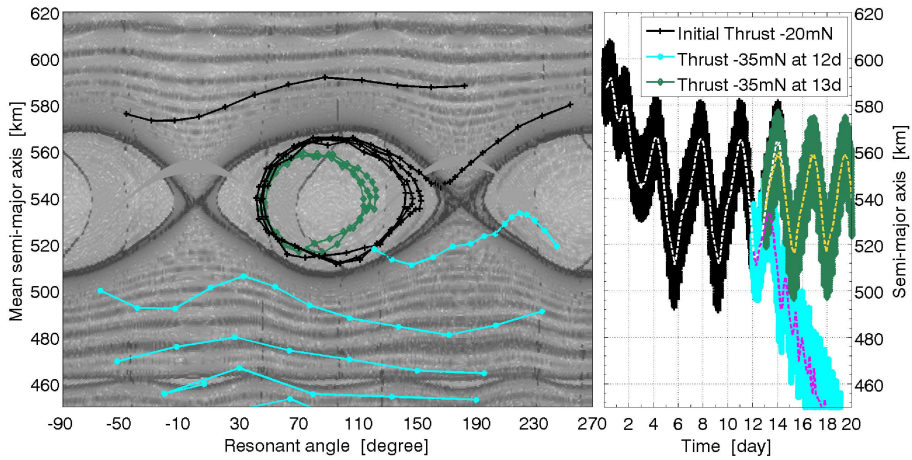}
    \caption{\label{ThrustEjection} 
      From a capture in the 1:1 resonance (black line), thrust of $35$mN at 
      two different times to try to escape from the trapping. 
      The force model contains a thrust and the gravitiy field until 
      degree and order 4. In right and left panels the dashed lines 
      (respectively the continued) represent 
      the evolution of the mean semi-major axis respectively with respect 
      to the time and to the resonant angle. In right panel we overlay the 
      evolution of the osculating semi-major axis. 
      The black line correspond to the dark line of the 
      Fig.~\ref{ThrustCapture}. The escape from the resonance 1:1 depends 
      on the phase of the resonant angle at the time of the operating 
      phase (changing of the thrust).}
  \end{center}
\end{figure} 
We note that the orbit with the change of thrust at $13$ days stays in the 
resonance 1:1 and is closer to the center of libration. The other case 
escapes from the resonance. The difference in the initial time of the $35$mN 
thrust induces a different value of the phase of the resonant angle at the 
time of the change of the thrust. These maps (the left panels of 
Figs.~\ref{ThrustEjection} and \ref{ThrustCapture}) 
allow us to easy understand that the probability of trapping or escape 
from the 1:1 resonance is dependent on the resonant angle phase.

%% \begin{figure}[htbp] 
%%   \begin{center}
%%     \includegraphics[draft=false,width=\textwidth]{tracerthrustdifsortireso2.eps}
%%     \caption{\label{ThrustSortie} ???.}
%%   \end{center}
%% \end{figure} 

%\newpage
\section{Conclusions}

The orbit dynamics of a polar space probe orbiting Vesta has been investigated. 
The proposed model includes the effects of the spherical harmonic 
approximation of the gravity field of the asteroid. We study 
the main ground-track resonances: 3:2, 4:3, 1:1, 2:3 and 1:2.  
We have developed a general method using an averaged Hamiltonian formulation. 
For each main ground-track resonance, we have analytically localized 
the resonance, determined the aperture and calculated the period at the 
stable equilibria. We have also discussed the effect of the error on the 
$C_{20}$ and $C_{22}$ coefficients on the properties of the 1:1 resonance. 

Our method allows to have an analytical global approach to search all 
main ground-track resonances and to compute the location
%periods of the free librations at the equilibria 
and the aperture of the resonances. 
The analytical results have been checked and validated 
numerically by performing numerical integrations of the complete systems. 

Our theory is able to reproduce, explain and complete the results 
of \cite{Tricarico2010}. 
The theory is general enough to be applied to a wide range of probes 
around any dwarf planet or most massive asteroid. In opposition to a 
numerical study, if the values of the spherical harmonic coefficients 
change (for example after new observations) our method stays valid. 

In a numerical way, we have found and justify that the increase 
of the eccentricity in the 2:3 resonance 
%1:1 and 2:3 are respectively due to the coefficient $J_{32}$ and $J_{22}$. 
is due to the $C_{22}$ coefficient. 
We have also numerically studied the probability of trapping 
in the 1:1 resonance that does not depend on the thrust if the initial 
semi-major axis of Dawn is upper than the 2:3 resonance location. 
We have also showed that the phase of the resonant angle is the key during 
the operating phase.

\section*{Acknowledgments}
%\begin{acknowledgements}
  Numerical simulations were made on the local computing ressources 
  ({\it Cluster URBM-SYSDYN}) at the University of Namur (FUNDP, Belgium).
%  Finally, the authors thank the two referees for their suggestions that 
%  contributed to improve this paper.
%\end{acknowledgements}

\end{document}